\pdfoutput=1

\documentclass[11pt,twoside,a4paper,cmspaper,final,collab]{cms-tdr}
\def\svnVersion{382111}\def\cmsCernNoTag{CERN-EP/2016-025}\def\cmsCernDate{\today}\def\cmsMessage{Published in the Journal of High Energy Physics as \href{http://dx.doi.org/10.1007/JHEP12(2016)088}{\doi{10.1007/JHEP12(2016)088.}}}

\begin{document}\cmsNoteHeader{EXO-14-004}

\hyphenation{had-ron-i-za-tion}
\hyphenation{cal-or-i-me-ter}
\hyphenation{de-vices}
\newcommand\mT{\rule{0pt}{2.3ex}}
\newcommand\mB{\rule[-1.2ex]{0pt}{0pt}}

\RCS$Revision: 369260 $
\RCS$HeadURL: svn+ssh://svn.cern.ch/reps/tdr2/papers/EXO-14-004/trunk/EXO-14-004.tex $
\RCS$Id: EXO-14-004.tex 369260 2016-09-29 19:04:40Z cpena $
\newlength\cmsFigWidth
\ifthenelse{\boolean{cms@external}}{\setlength\cmsFigWidth{0.85\columnwidth}}{\setlength\cmsFigWidth{0.4\textwidth}}
\ifthenelse{\boolean{cms@external}}{\providecommand{\cmsLeft}{top}}{\providecommand{\cmsLeft}{left}}
\ifthenelse{\boolean{cms@external}}{\providecommand{\cmsRight}{bottom}}{\providecommand{\cmsRight}{right}}
\providecommand{\MR}{\ensuremath{M_R}\xspace}
\providecommand{\RR}{\ensuremath{R}\xspace}
\providecommand{\NA}{\text{---}\xspace}
\newlength\cmsTabSkip\setlength{\cmsTabSkip}{1.5ex}
\newcolumntype{x}{D{,}{\,\pm\,}{-1}}
\newcommand*{\vv}[1]{\ensuremath{\vec{#1\mkern0mu}}}

\cmsNoteHeader{EXO-14-004}
\title{Search for dark matter particles in proton-proton collisions at $\sqrt{s} = 8$\TeV
  using the razor variables}

\date{\today}

\abstract{A search for dark matter particles directly produced in proton-proton collisions recorded by the CMS experiment at the LHC is
presented.
  The data correspond to an integrated luminosity of 18.8\fbinv, at a
  center-of-mass energy of 8\TeV. The event selection requires at least two
  jets and no isolated leptons. The razor variables are
  used to quantify the transverse momentum balance in the jet
  momenta. The study is performed separately for events with and
  without jets originating from b quarks.
 The observed yields are consistent with the expected
 backgrounds and, depending on the nature of the production mechanism,
dark matter production at the LHC is excluded at 90\% confidence level
for a mediator mass scale $\Lambda$ below 1\TeV.
The use of razor variables yields results that complement those previously published.
}
\hypersetup{%
 pdfauthor={CMS Collaboration},%
 pdftitle={Search for dark matter particles in proton-proton collisions at
   sqrt(s) = 8 TeV using the razor variables},%
 pdfsubject={CMS},%
 pdfkeywords={CMS, physics, razor, SUSY, btag}}

\maketitle
\section{Introduction}

The existence of dark matter (DM) in the universe, originally proposed~\cite{Zwicky:1937zza} to
reconcile observations of the Coma galaxy cluster with
the prediction from the virial theorem, is
commonly accepted as the explanation of many experimental
phenomena in astrophysics and cosmology, such as galaxy rotation
curves~\cite{vandeHulst,Rubin:1980zd}, large structure
formation~\cite{White:1987yr,Carlberg:1989yr,Springel:2005nw}, and the
observed
spectrum~\cite{Smoot:1992td,deBernardis:2000gy,Spergel:2006hy,Ade:2013zuv}
of the cosmic microwave background~\cite{Bardeen:1985tr}. A global fit to
cosmological data in the $\Lambda$CDM model (also known as
the standard model of cosmology)~\cite{Cen:2000xv} suggests that
approximately 85\% of the mass of the universe is attributable to
DM~\cite{Ade:2013zuv}. To accommodate these observations and the
dynamics of colliding galaxy clusters~\cite{Clowe:2006eq}, it has been
hypothesized that DM is made mostly of weakly
interacting massive particles
(WIMPs), sufficiently massive to be in nonrelativistic motion
following their decoupling from the hot particle plasma in the early
stages of the expansion of the universe.

While the standard model (SM) of particle physics does not include a
viable DM candidate, several models of physics beyond the SM, e.g.,
supersymmetry (SUSY)~\cite{Ramond,Golfand,Volkov,Wess,Fayet} with $R$-parity
conservation, can accommodate the existence of WIMPs. In these models,
pairs of DM particles can be produced in proton-proton (pp) collisions at
the CERN LHC. Dark matter particles would not leave a detectable signal in
a particle detector. When produced in association with high-energy
quarks or gluons, they could provide event topologies with
jets and a transverse momentum (\pt) imbalance ($\ptvecmiss$). The magnitude of $\ptvecmiss$ is referred to as missing transverse energy ($\ETm$).
The ATLAS and CMS collaborations have reported searches for events with one
high-\pt jet and large $\ETm$~\cite{Aad:2011xw,Chatrchyan:2012me}, which are sensitive to such topologies.
In this paper, we refer to these studies as monojet searches. Complementary studies of events with
high-\pt photons~\cite{Khachatryan:2014rwa,Aad:2014tda}; \PW,
\cPZ,~or
Higgs~bosons~\cite{Aad:2013oja,Aad:2014vka,Aad:2015dva,Aad:2015yga};
b jets~\cite{Aad:2014vea} and top quarks~\cite{Aad:2014vea,CMS:b2g12-022,CMS:semilepTop}; and leptons~\cite{ATLAS:2014wra,Khachatryan:2014tva}
have also been performed.

This paper describes a search for dark matter particles $\chi$ in events with at least two jets of comparable transverse momenta
and sizable $\ETm$. The search is based
on the razor variables $\MR$ and $\RR^2$~\cite{rogan,razor2010}.  Given a
dijet event, these variables are computed from the two jet momenta $\vv{p}^{j_1}$ and
$\vv{p}^{j_2}$, according to the following
definition:
 \begin{align}
 \begin{split}
 \MR  & =
 \sqrt{
  (\abs{\vv{p}^{j_1}}+\abs{\vv{p}^{j_{2}})^2 -
  (p^{j_1}_{z}+p^{j_2}_{z})^2}},
\\
 R  & =   \frac{M^{\RR}_\mathrm{T}}{\MR},
\label{eq:razor}
\end{split}\\
\intertext{with}
 \label{eq:MTR}
 M^{\RR}_\mathrm{T}  &=   \sqrt{ \frac{\ETm(\pt^{j_1}+\pt^{j_2}) -
   \ptvecmiss {\cdot}
   (\vv{p}_{\mathrm{T}}^{\,j_1}+\vv{p}_{\mathrm{T}}^{j_2})}{2}}.
\end{align}

\begin{figure}
 \centering
\includegraphics[width=0.35\textwidth]{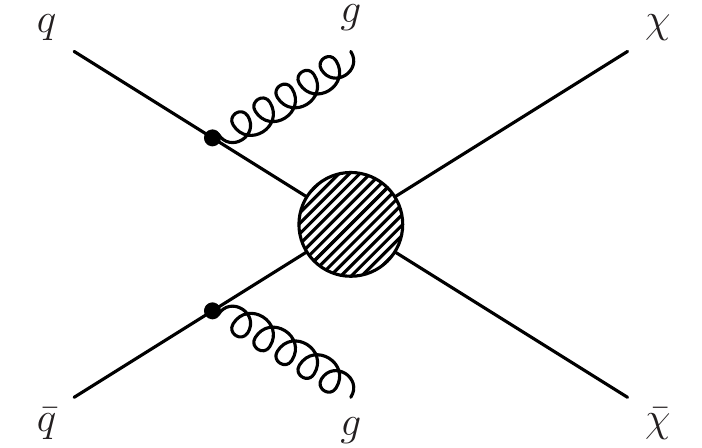}
\includegraphics[width=0.35\textwidth]{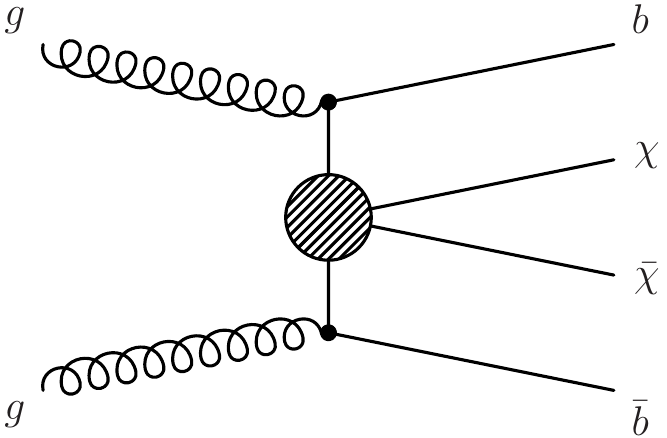}
 \caption{Feynman diagrams for the pair production of DM particles
   corresponding to an effective field theory using a vector or
   axial-vector operator (left), and a scalar operator (right).\label{fig:DMdiamgrams}}
\end{figure}

In the context of SUSY, $\MR$ provides an estimate of the
underlying mass scale of the event, and quantity $M^{\RR}_\mathrm{T}$ is a transverse
observable that includes information about the
topology of the event. The variable $\RR^2$ is designed to reduce QCD
multijet background; it is correlated with the angle
between the two jets, where co-linear jets have large $\RR^2$ while back-to-back jets have small $\RR^2$.
These variables have been used to study the production of non-interacting
particles in cascade decays of heavier partners, such as squarks and
gluinos in SUSY models with $R$-parity
conservation~\cite{Chatrchyan:2014goa,Razor8TeV}. The sensitivity of
these variables to direct DM production was suggested in
Ref.~\cite{Fox:2012ee}, where it was pointed out that the dijet event
topology provides good discrimination against background processes,
with a looser event selection than that applied in the monojet searches.
Sensitivity to DM production is most enhanced for large values of
$\RR^2$, while categorizing events based on the value of $\MR$
improves signal to background discrimination and yields significantly improved
search sensitivity to a broader and more inclusive class of DM models.
The resulting sensitivity is expected to be
comparable to that of monojet searches~\cite{Fox:2012ee,Papucci:2014iwa}. This
strategy also offers the possibility to search for DM particles that
couple preferentially to b quarks~\cite{Agrawal:2014una}, as proposed
to accommodate the observed excess of photons with energies between
1 and 4\GeV in the gamma ray spectrum of the galactic center data collected by the Fermi-LAT gamma-ray space
telescope~\cite{Hooper:2010mq}. The results are interpreted using an
effective field theory approach and the Feynman diagrams for DM pair production are shown in Fig.~\ref{fig:DMdiamgrams}.

Unlike the SUSY razor searches~\cite{Razor8TeV,razor2010}, which focus on events with large
values of $\MR$, this study also considers events with small values
of $\MR$, using $\RR^2$ to discriminate
between signal and background, in a kinematic region ($\RR^2 >
0.5$) excluded by the baseline selection of Refs.~\cite{Razor8TeV,razor2010}.

A data sample corresponding to an integrated luminosity of
18.8\fbinv of pp collisions at a center-of-mass energy of 8\TeV  was collected by the CMS experiment
 with a trigger based on a
loose selection on $\MR$ and $\RR^2$. This and other
special triggers were operated in 2012 to record events at a rate
higher than the CMS computing system could process during data
taking. The events from these triggers were stored on tape and their
reconstruction was delayed until 2013, to profit from the larger availability of processing resources during the LHC shutdown.
These data, referred to as ``parked data''~\cite{CMS-DP-2012-022},
enabled the exploration of events with small $\MR$
values, thereby enhancing the sensitivity to direct DM production.

This paper is organized as follows: the CMS detector is briefly described
in Section~\ref{cmsdetector}.  Section~\ref{sec:sample}
describes the data and simulated samples of events used in the
analysis. Sections~\ref{sec:selection} and~\ref{sec:sampleDef} discuss
the event selections and categorization, respectively. The
estimation of the background is described in Section~\ref{sec:bkg}.
The systematic uncertainties are discussed in Section~\ref{sec:sys},
while Section~\ref{sec:interpretation} presents the results and the
implications for several models of DM production. A summary is given in Section~\ref{sec:conclusions}.

\section{The CMS detector}\label{cmsdetector}

The central feature of the CMS apparatus is a superconducting solenoid
of 6\unit{m} internal diameter, providing a magnetic field of
3.8\unit{T}. Within the solenoid volume are a silicon
pixel and strip tracker, a lead tungstate crystal electromagnetic
calorimeter (ECAL), and a brass and scintillator hadron calorimeter
(HCAL), each composed of a barrel and two endcap sections.
When combining information from the entire detector, the jet energy resolution amounts typically to 15\% at 10\GeV, 8\% at 100\GeV, and 4\% at 1\TeV~\cite{Chatrchyan:2013dga}. Muons are
measured in gas-ionization detectors embedded in the steel flux-return
yoke outside the solenoid. Forward calorimeters extend the
pseudorapidity ($\eta$)~\cite{Chatrchyan:2008zzk} coverage provided by the
barrel and endcap detectors. The first level (L1) of the CMS trigger system, composed of custom
hardware processors, uses information from the calorimeters and muon
detectors to select the most interesting events in a fixed time
interval of less than 4\mus. The high-level trigger (HLT) processor
farm further decreases the event rate from around 100\unit{kHz} to
around 400\unit{Hz}, before data storage. A more
detailed description of the CMS detector, together with a definition
of the coordinate system used and the basic kinematic variables, can
be found in Ref.~\cite{Chatrchyan:2008zzk}.

\section{Data set and simulated samples}
\label{sec:sample}
The analysis is performed on events with two jets reconstructed at L1
in the central part of the detector ($\abs{\eta}< 3.0$). The L1 jet
triggers are based on the sums of
transverse energy in regions $\Delta\eta\times\Delta\phi$
approximately 1.05$\times$1.05 in
size~\cite{Chatrchyan:2008zzk} (where $\phi$ is the azimuthal angle in the plane transverse to the LHC beams.).
At the HLT, energy deposits in ECAL and HCAL are clustered into jets and the
razor variables $\RR^2$ and $\MR$ are computed. In the
HLT, jets are defined using the {\FASTJET}~\cite{fastjet}
implementation of the anti-\kt~\cite{antikt} algorithm, with
a distance parameter equal to 0.5. Events with at least two jets
with $\pt>64$\GeV are
considered. Events are selected with $\RR^2> 0.09$ and
$\RR^2 \times \MR > 45$\GeV. This selection
rejects the majority of the background, which tends to have low $\RR^2$ and low
$\MR$ values, while keeping the events in the signal-sensitive
regions of the ($\MR$, $\RR^2$) plane.
The trigger efficiency, measured using a pre-scaled trigger with very loose thresholds, is
shown in Table~\ref{tab:Trigger}.  The requirements described above correspond to the least stringent event selection, given the constraints on the maximum acceptable rate.

\begin{table}[htb]
\centering
  \topcaption{\label{tab:Trigger}Measured trigger efficiency for different
    $\MR$ regions. The selection $\RR^2 > 0.35$ is applied. The uncertainty
    shown represents the statistical uncertainty in the measured efficiency.
}
\begin{tabular}{lccc}
  \hline
  $\MR$ region (\GeVns{}) & 200--300 &  300--400 &
  400--3500 \rule{0pt}{2.3ex} \rule[-1.2ex]{0pt}{0pt}\\
  \hline
  Trigger efficiency (\%) & $91.1\pm ^{1.5}_{1.7}$ &
  $90.7\pm^{2.3}_{2.9}$ & $94.4 \pm ^{2.4}_{3.6}$ \rule{0pt}{2.3ex} \rule[-1.2ex]{0pt}{0pt}\\
  \hline
\end{tabular}
\end{table}

Monte Carlo (MC) simulated signal and background samples are generated with the
leading order matrix element generator {\MADGRAPH
  v5.1.3}~\cite{Alwall:2011uj,Alwall:2014hca} and the CTEQ6L parton
distribution function set~\cite{Pumplin:2002vw}. The generation
includes the \PYTHIA 6.4.26~\cite{Sjostrand:2006za} Z2* tune, which is
derived from Z1 tune~\cite{Field:2010bc} based on the CTEQ5L set.
Parton shower and hadronization effects are included by matching the generated events to \PYTHIA, using the MLM matching algorithm~\cite{Hoche:2006ph}.
The events are processed with a \GEANTfour~\cite{G4} description of the CMS apparatus to include
detector effects. The simulation samples for SM
background processes are scaled to the
integrated luminosity of the data sample (18.8\fbinv), using
calculations of the inclusive production cross sections at the next-to-next-to-leading
order (NNLO) in the perturbative QCD
expansion~\cite{WatNNLO,ZatNNLO,TTbaratNNLO}.
 The signal processes corresponding to pair production of DM particles
 are simulated with up to two additional partons with $\pt>80$\GeV.

\section{Event selection}\label{sec:selection}

Events are selected with at least one reconstructed
interaction vertex within $\abs{z}<24$\cm. If more than one vertex is
found, the one with the highest sum of the associated track momenta squared is used as the
interaction point for event reconstruction. Events containing
calorimeter noise, or large missing transverse momentum
due to beam halo and instrumental effects (such as jets near
non-functioning channels in the ECAL) are removed from the analysis~\cite{MET_8TeV}.

A particle-flow (PF) algorithm~\cite{PF1,CMS-PAS-PFT-10-001} is used to reconstruct
and identify individual particles with an optimized combination of
information from the various elements of the CMS detector. The energy
of photons is directly obtained from the ECAL measurement, corrected
for zero-suppression effects. The energy of electrons is determined
from a combination of the electron momentum at the primary interaction
vertex as measured by the tracker, the energy of the corresponding
ECAL cluster, and the energy sum of all bremsstrahlung photons (or emissions)
spatially compatible with originating from the electron track. The
energy of muons is obtained from the curvature of the associated
track. The energy of charged hadrons is determined from a combination
of their momentum measured in the tracker and the matching ECAL and
HCAL energy deposits, corrected for zero-suppression effects and for
the response function of the calorimeters to hadronic
showers. Finally, the energy of neutral hadrons is obtained from the
corresponding corrected ECAL and HCAL energies. Contamination of the
energy determinations from other pp collisions is mitigated by
discarding the charged PF candidates incompatible with originating from the main vertex.  Additional
energy from neutral particles is subtracted on average when computing
lepton (electron or muon) isolation and jet energy. This contribution is estimated as the
per-event energy deposit per unit area, in the cone $\Delta R = \sqrt{\smash[b]{(\Delta
  \eta)^2+(\Delta\phi)^2}}=0.3$, times the considered jet size or
isolation cone area.

To separate signal from the main backgrounds it is necessary to
identify electrons (muons) with $\pt>15$\GeV and  $\abs{\eta}<2.5$
(2.4). In order to reduce the rate for misidentifying hadrons as 
leptons, additional requirements based on the
quality of track reconstruction and isolation are applied. Lepton isolation is
defined as the scalar \pt sum of all PF candidates other than the
lepton itself, within a cone of size $\Delta R = 0.3$, and normalized to the lepton \pt. A
candidate is identified as a lepton if the isolation variable is found to be smaller than 15\%.
For electrons~\cite{ElectronsCMS}, a characteristic of the shower
shape of the energy deposit in the ECAL (the shower width in the
$\eta$ direction) is used to further reduce the contamination from hadrons.
PF candidates with $\pt > 10$\GeV that are not consistent with muons and satisfy the same isolation
requirements as those used for electrons are also identified to increase the
lepton selection efficiency as well as to identify single-prong tau decays.

Jets are formed by clustering the PF candidates, using the anti-\kt algorithm with distance
parameter 0.5. Jet momentum is determined as the vectorial sum of all
particle momenta in the jet, and is found from simulation to be within
5\% to 10\% of the generated hadron level jet momentum over the whole \pt spectrum and
detector acceptance. Jet energy corrections
are derived from simulation, and are confirmed with in situ
measurements of the energy balance in dijet and photon+jet events.
Any jet whose momentum points within a cone of $\Delta R < 0.3$ around any identified
electron, muon, or isolated track is discarded.
Additional selection criteria are applied to each event to remove
spurious jet-like features originating from isolated noise patterns in certain HCAL regions.
We select events containing at least two jets with $\pt>80$\GeV and $\abs{\eta}<2.4$, for
which the corresponding L1 and HLT requirements are maximally
efficient. The combined secondary vertex (CSV) b-tagging
algorithm~\cite{btag8TeV,btag7TeV} is used to identify jets originating from b
quarks. The loose and tight working points of the CSV algorithm, with
85\% (10\%) and 50\% (0.1\%) identification efficiency (misidentification probability) respectively, are
used to assign the selected events to categories based on the number
of b-tagged jets, as described below.

In order to compute the razor variables inclusively, the event is forced into a two-jet topology, by forming two \textit{megajets}~\cite{Chatrchyan:2014goa} out of all the reconstructed
jets with $\pt>40$\GeV and $\abs{\eta}<2.4$. All possible assignments of
jets to the megajets are considered, with the requirement that a
megajet consist of at least one jet. The sum of the four-momenta of
the jets assigned to a megajet defines the megajet
four-momentum. When more than two jets are reconstructed, more than
one megajet assignment is possible. We select the assignment that
minimizes the sum of the invariant masses of the two megajets.
In order to reduce the contamination from multijet production, events are
rejected if the angle between the two selected megajets in the
transverse plane $\abs{ \Delta\phi (j_{1}, j_{2})}$ is larger
than 2.5 radians. The momenta of the two megajets are used to compute
the razor variables, according to Eq.~(\ref{eq:razor},~\ref{eq:MTR}).  Events are
required to have $\MR>200$\GeV and $\RR^2>0.5$.

\section{Analysis Strategy}\label{sec:sampleDef}

To enhance the DM signal and suppress background contributions from the $\PW$+jets and $\ttbar$ processes,
we veto events with selected electrons, muons, or isolated charged PF candidates.
We define three different search regions based on the number of b-tagged jets.
The zero b-tag search region contains events where no jets were identified with the CSV loose
b-tagging criterion; the one b-tag search region contains events where exactly one jet
passed the CSV tight criterion; and the two b-tag search region contains events where two or more
jets passed the CSV tight criterion. Events in the zero b-tag search region are further classified
into four categories based on the value of $\MR$, to enhance signal to background
discrimination for a broad class of DM models:
(i) \textit{very low} $\MR$ (VL), defined by $200<\MR \leq 300$\GeV;
(ii) \textit{low} $\MR$ (L), with $300 <\MR \leq 400$\GeV;
(iii) \textit{high} $\MR$ (H), with $400 <\MR\leq 600$\GeV;
and (iv) \textit{very high} $\MR$ (VH), including events with $\MR>600$\GeV.
Because of the limited size of the data sample, no further
categorization based on $\MR$ is made for the one and two b-tag search regions.
Within each category, the search is performed in bins of the $\RR^2$
variable, with the binning chosen such that the expected background yield
in each bin is larger than one event, as estimated from Monte Carlo simulation.

In the H and VH categories, 3\% and 35\% respectively of the selected
events were also selected in the monojet search~\cite{monojet8TeV}, which used data from
the same running period. The overlap in the L and VL categories is negligible, while the
overlapping events in the H and VH categories were shown not to have an impact on the final
sensitivity. Consequently, the results from this analysis and from the monojet analysis
are largely statistically independent.

The main backgrounds in the zero b-tag search region are from the $\PW(\ell\nu)$+jets
and $\cPZ(\PGn\PAGn)$+jets processes, while the dominant background in the one and two
b-tag search regions is the $\ttbar$ process. To estimate the contribution of these
backgrounds in the search regions, we use a data-driven method that extrapolates
from appropriately selected control regions to the search region, assisted by
Monte Carlo simulation. A detailed description of the background
estimation method is discussed in Section~\ref{sec:bkg}.

To estimate the $\PW(\ell\nu)$+jets and $\cPZ(\PGn\PAGn)$+jets background in the
zero b-tag search region, we define the 1$\mu$ control region by selecting events
using identical requirements to those used in the search region, with the exception
of additionally requiring one selected muon. Events in this control region are extrapolated
to the search region in order to estimate the background. In addition, we define
the 2$\mu$ control region, enhanced in the $\cPZ$+jets process, by requiring two selected
muons with invariant mass between $80$\GeV and $100$\GeV. The 2$\mu$ control region is used to perform a cross-check prediction for the
1$\mu$ control region, and the systematic uncertainties in
background prediction are estimated based on this comparison.

To estimate the $\ttbar$ background in the one and two b-tag search regions,
we define the 1$\mu$b and 2$\mu$b control regions, by requiring at least one
jet satisfying the CSV tight b-tagging criterion along with one and two selected
muons respectively. Both of these control regions are dominated by the
$\ttbar$ process. The $\ttbar$ background prediction is estimated by extrapolating
from the 2$\mu$b control region, while the 1$\mu$b control region is used
as a cross-check to estimate systematic uncertainties. Finally, we define
the $\cPZ(\mu\mu)$b control region by requiring two muons with invariant
mass between $80$\GeV and $100$\GeV. This is used to estimate the
$\cPZ(\PGn\PAGn)$+jets background in the one and two b-tag search regions.

The definitions of the search and control regions, and their use in this analysis are
summarized in Tables~\ref{tab:boxes}~and~\ref{tab:boxes1}.

\begin{table}
  \topcaption{\label{tab:boxes} Analysis regions for
    events with zero identified b-tagged jets. The definition of these
    regions is based on the muon multiplicity, the output of the CSV b-tagging
    algorithm, and the value of $\MR$. For all the regions,
    $\RR^2>0.5$ is required.}
  \centering
 \begin{tabular}{llll}
  \hline
  \multicolumn{1}{c}{analysis region}  & \multicolumn{1}{c}{purpose} &  \multicolumn{1}{c}{b-tagging selection}  &  \multicolumn{1}{c}{$\MR$ category} \\
  \hline
  \multirow{2}{*}{0$\mu$}  & \multirow{2}{*}{signal search region} &   &  \\
   &   &  & $200<\MR \leq 300$\GeV (VL)\\
\multirow{2}{*}{1$\mu$}  &  \multirow{2}{*}{$\PW(\ell\nu)$ control region} & \multirow{2}{*}{no CSV loose jet} &$300<\MR \leq 400$\GeV (L) \\
   &   &  &  $400<\MR \leq 600$\GeV (H)\\
\multirow{2}{*}{2$\mu$}  &  \multirow{2}{*}{$\cPZ(\ell \ell)$ control region} &  & \phantom{$400<$}$\MR > 600$\GeV (VH)\\
&   &  & \\
\hline
\end{tabular}
\end{table}

\begin{table}
  \topcaption{\label{tab:boxes1} Analysis regions for
    events with identified b-tagged jets. The definition of these
    regions is based on the muon multiplicity, the output of the CSV b-tagging
    algorithm, and the value of $\MR$. For all the regions,
    $\RR^2>0.5$ is required.}
  \centering
 \begin{tabular}{llll}
\hline
\hline
 \multicolumn{1}{c}{analysis region}  & \multicolumn{1}{c}{purpose}  &  \multicolumn{1}{c}{b-tagging
                                      selection}  &
                                                    \multicolumn{1}{c}{$\MR$
                                                    category} \\
\hline
 0$\mu$bb  & \multirow{2}{*}{signal serach region} &$\geq$2 CSV tight jets  &  \multicolumn{1}{r}{\multirow{8}{*}{$\MR > 200$\GeV}} \\
0$\mu$b  & & $= 1$ CSV tight jet &  \\\\
1$\mu$b  & $\ttbar$ control region &  \multirow{2}{*}{$\geq$1 CSV tight jets}  & \\
2$\mu$b  & $\ttbar$ control region &   &  \\\\
$\cPZ(\mu\mu)$b & $\cPZ(\ell \ell)$ control region  &$\geq$1 CSV loose jets  &  \\
\hline
\end{tabular}
\end{table}

\section{Background estimation}\label{sec:bkg}

The largest background contribution to the zero b-tag search region is from events in which a W or Z boson is
produced, in association with jets, decaying to final states with one or more neutrinos. These
background processes are referred to as $\PW(\ell\nu)$+jets
and $\cPZ(\PGn\PAGn)$+jets events. Additional backgrounds arise from events involving the production of top quark pairs, and from
events in which a $\cPZ$ boson decays to a pair of charged
leptons. These processes are referred to as $\ttbar$ and
$\cPZ(\ell \ell)$+jets, respectively. Using
simulated samples, the contribution from other SM processes, such as
diboson and single top production, is found to be negligible.

The main background in the one and two b-tag search regions comes from $\ttbar$ events.
The use of the tight working point of the CSV
algorithm reduces the $\cPZ(\PGn\PAGn)$+jets and $\PW(\ell\nu)$+jets contribution as
shown in Table~\ref{tab:bkg0mu}. Multijet production, which is the most abundant source of events with jets and
unbalanced \pt, contributes to the search region primarily due to
instrumental mismeasurement of the energy of jets. As a result the
\MET direction tends to be highly aligned in the azimuthal coordinate with
the razor megajets. The requirement on the razor variables and
$\abs{ \Delta\phi (j_{1}, j_{2})}$ reduces the multijet
background to a negligible level, which is confirmed by checking
data control regions with looser cuts on the razor variables.

\subsection{Background estimation for the zero b-tag search region}
\label{sec:bkgzmu}

To predict the background from $\PW(\ell\nu)$+jets and $\cPZ(\PGn\PAGn)$+jets in
the zero b-tag search region, we use a data-driven method that extrapolates
the observed data yields in the 1$\mu$ control region to the search region.
Similarly, the observed yield in the 2$\mu$ control region allows the estimation of
the contribution from $\cPZ(\ell \ell)$+jets background process. Each
$\MR$ category is binned in $\RR^2$. Events in which the $\PW$ or $\cPZ$ boson
decayed to muons are used to extrapolate to cases where they decay to electrons
or taus. 

The background expected from $\PW$ and $\cPZ$ boson production, in
each $\RR^2$ bin and in each $\MR$ category of the 0$\mu$
sample, is computed as
\begin{equation}
  n_{i}^{0\mu} =  \Bigl(n_{i}^{1\mu} - N_{i}^{\ttbar,1\mu} - N_{i}^{\cPZ(\ell
    \ell)+\text{jets},1\mu}\Bigr) \frac{N_{i}^{\PW(\ell\nu)+\text{jets},0\mu}+N_{i}^{\cPZ(\PGn\PAGn)+\text{jets},0\mu}}{N_{i}^{\PW(\ell\nu)+\text{jets},1\mu}} +
\Bigl(n_{i}^{2\mu} - N_{i}^{\ttbar,2\mu}\Bigr) \frac{N_{i}^{\cPZ(\ell \ell)+\text{jets},0\mu}}{N_{i}^{\cPZ(\ell \ell)+\text{jets},2\mu}},
\label{eqn:pred}
\end{equation}
where $n_{i}^{k\mu}$ labels the data yield in bin $i$ for the sample
with $k$ muons, and $N_{i}^{X,k\mu}$ indicates the corresponding yield
for process $X$, derived from simulations. This background estimation method relies on
the assumption that the kinematic properties of events in which $\PW$ and $\cPZ$ bosons are
produced are similar.

To estimate the accuracy of the background estimation method, we perform a cross-check
by predicting the background in the 1$\mu$ control region using the observed data yield
in the 2$\mu$ control region. The Monte Carlo simulation is used to perform this extrapolation
analogous to the calculation in Equation~\ref{eqn:pred}.
The small contribution from the $\ttbar$ background process is also estimated using the simulated
samples. In Tables~\ref{tab:bkg1mu}~and~\ref{tab:bkg2mu}, the observed yields in the 1$\mu$ and 2$\mu$ control regions
respectively are compared to the estimate derived from data. In Tables~\ref{tab:bkg1mu}-\ref{tab:bkg0muWITHB},
the contribution of each process as predicted directly by simulated samples are also given.

\begin{table}[htb]
  \topcaption{\label{tab:bkg1mu} Comparison of the observed yield in the
    1$\mu$ control region in each $\MR$ category and the
    corresponding data-driven background estimate obtained by extrapolating from the 2$\mu$ control region. The uncertainty in
    the estimates takes into account both the statistical
    and systematic components. The contribution of each individual background process is also shown, as
estimated from simulated samples, as well as the total MC predicted yield.}
\centering
\resizebox{\textwidth}{!}{
 \begin{tabular}{c*{6}{x}r}
   \hline
 \multicolumn{1}{c}{$\MR$ category} &  \multicolumn{1}{c}{$\cPZ(\PGn\PAGn)$+jets}  &  \multicolumn{1}{c}{$\PW(\ell
 \nu)$+jets}  &  \multicolumn{1}{c}{$\cPZ(\ell \ell)$+jets}  &  \multicolumn{1}{c}{$\ttbar$}  & \multicolumn{1}{c}{MC predicted}
 &\multicolumn{1}{c}{Estimated}  &  \multicolumn{1}{c}{Observed} \mT \mB\\
   \hline
   VL  &  0.7,0.3 & 4558,32 & 133,3 & 799,9 & 5491,33 & 5288,511 & 5926\mT\\
   L  &  0.5,0.3 & 1805,17 & 44,2 & 213,4 & 2063,18 & 1840,233 & 2110 \\
   H  &  0.1,0.1 & 915,11 & 16,1 & 66,2 & 997,11 & 629,240 & 923 \\
   VH  &   \multicolumn{1}{c}{$<$0.1}  & 183,5 & 2.6,0.2 & 8.5,0.8 & 194,5 & 166,93 & 143\mB\\
   \hline
 \end{tabular}
}
\end{table}

\begin{table}[htb]
  \centering
  \topcaption{\label{tab:bkg2mu} Comparison of the observed yield for
    the 2$\mu$ control region in each $\MR$ category and the
    corresponding prediction from background simulation. The quoted uncertainty
    in the prediction reflects only the size of the simulated sample. The
    contribution of each individual background process is also shown, as
    estimated from simulated samples.}
 \begin{tabular}{*{6}{c}r}
   \hline
$\MR$  category&  $\cPZ(\PGn\PAGn)$+jets  &  $\PW(\ell
   \nu)$+jets  &  $\cPZ(\ell \ell)$+jets  &  $\ttbar$  &  MC predicted
   &  \multicolumn{1}{c}{Observed} \mT\mB\\
   \hline
   VL  &   $<$0.1  &  $<$0.1  & $214\pm4$ & $1.9\pm0.3$ & $215\pm4$ & 207\mT\\
   L  &    $<$0.1 & $0.4\pm0.3$ & $88\pm2$ & $0.5\pm0.2$ & $89\pm2$ & 78 \\
   H  &   $<$0.1  & $0.1\pm0.1$ & $48\pm1$ & $0.1\pm0.1$ & $48\pm1$ & 30 \\
   VH  &   $<$0.1  &  $<$0.1  & $10\pm1$ & $0.1\pm0.1$ & $10\pm1$ & 7\mB\\
   \hline
\end{tabular}
\end{table}

\begin{figure}
 \centering
   \includegraphics[width=0.48\textwidth]{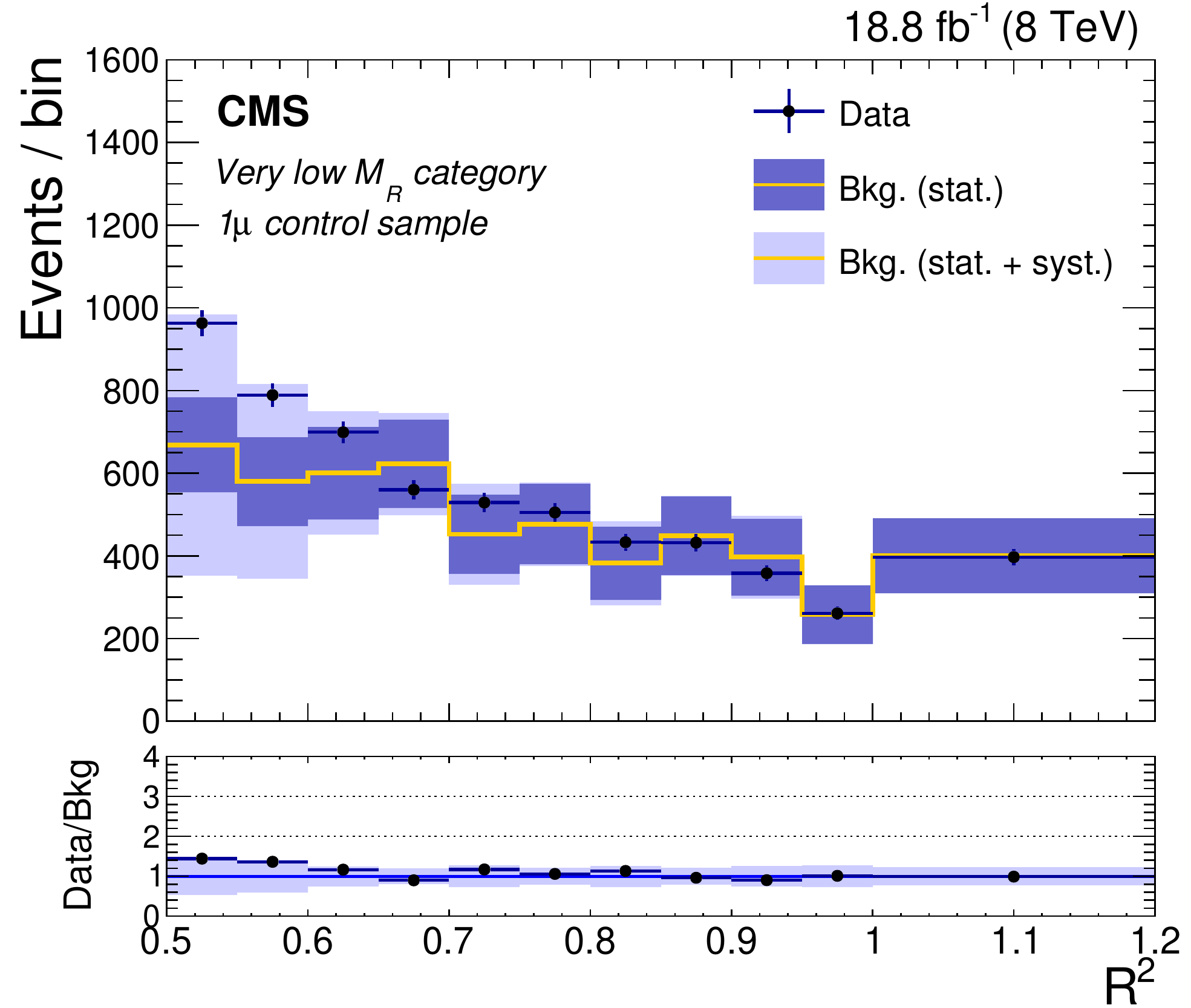}
   \includegraphics[width=0.48\textwidth]{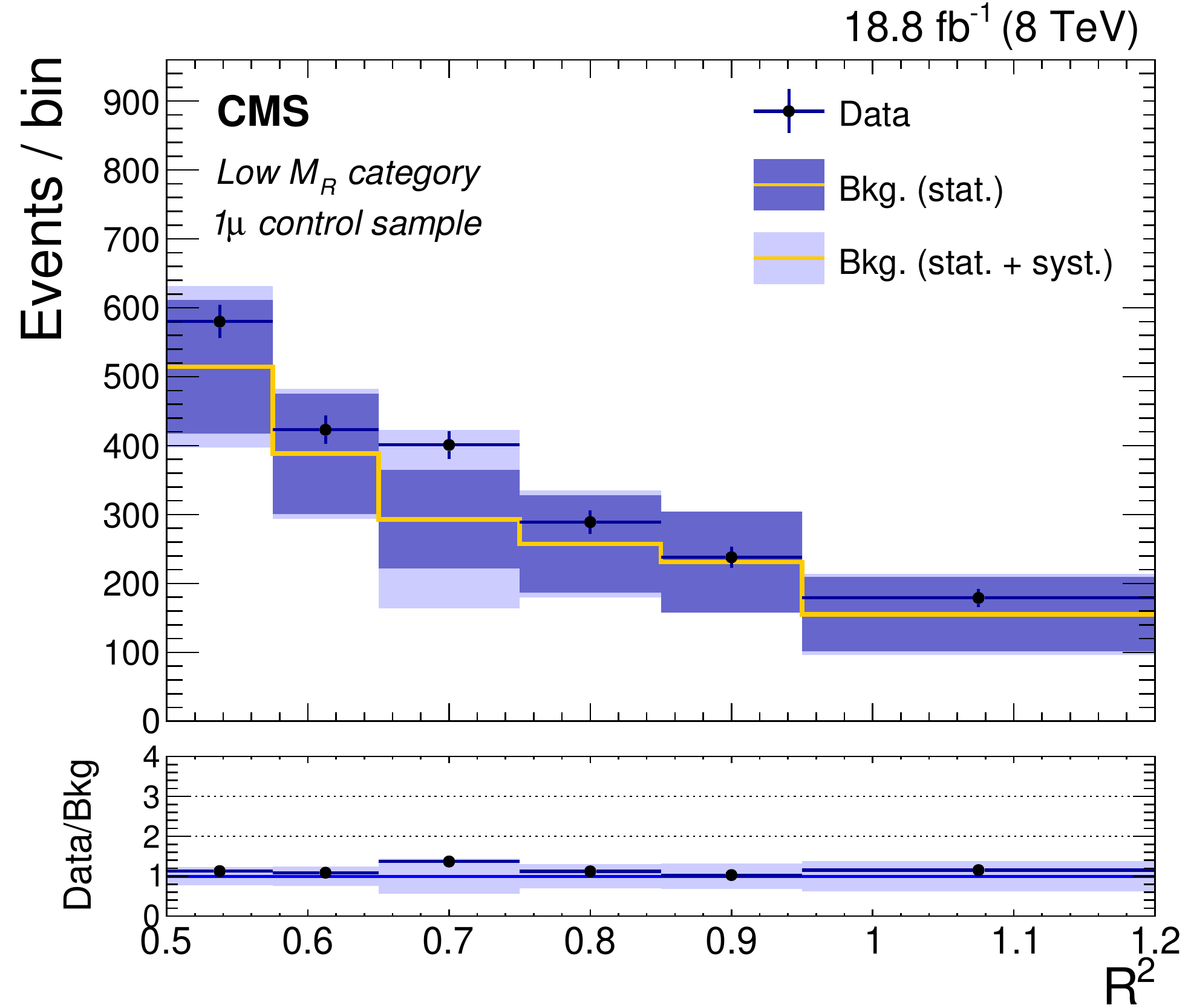}
   \includegraphics[width=0.48\textwidth]{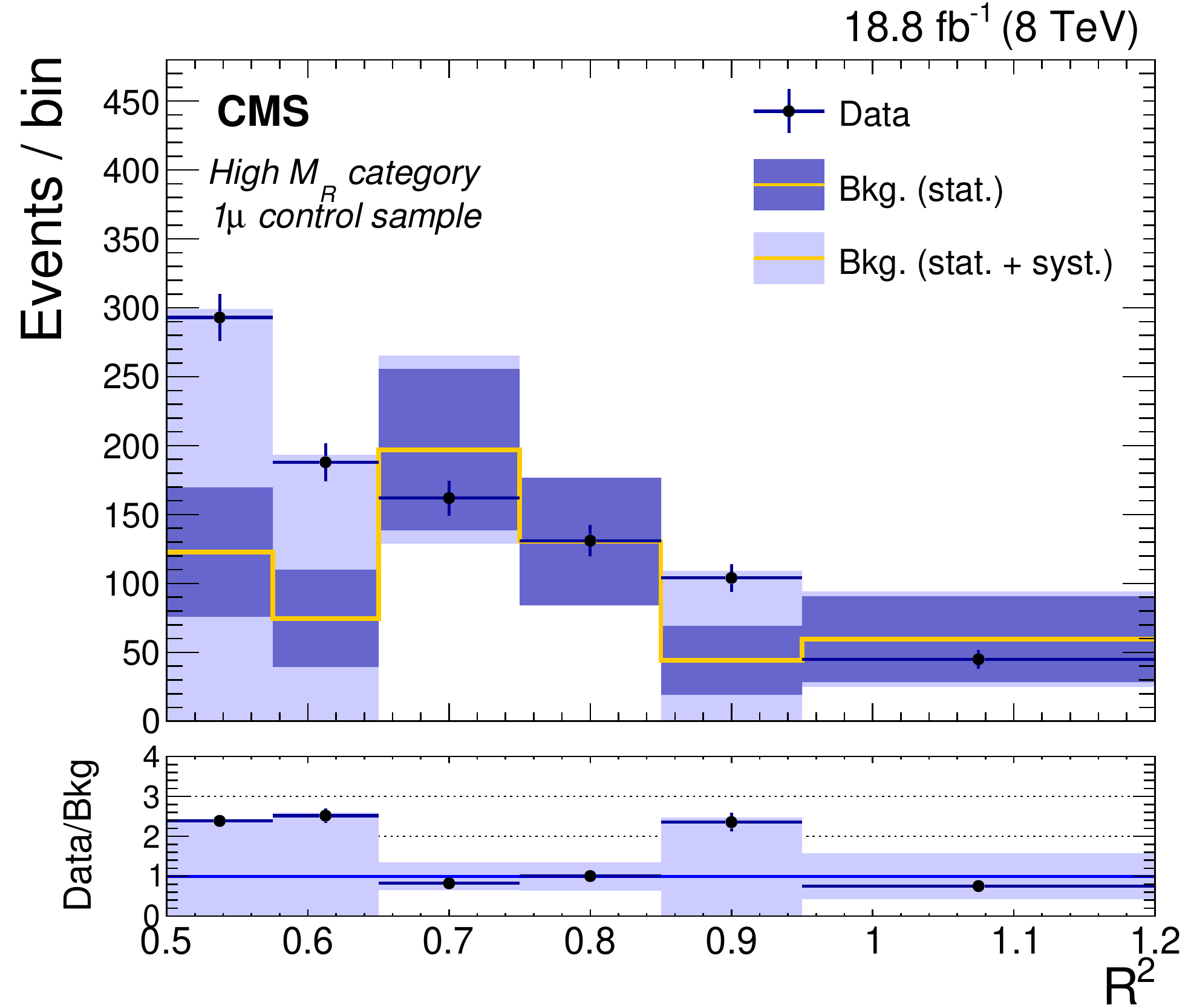}
   \includegraphics[width=0.48\textwidth]{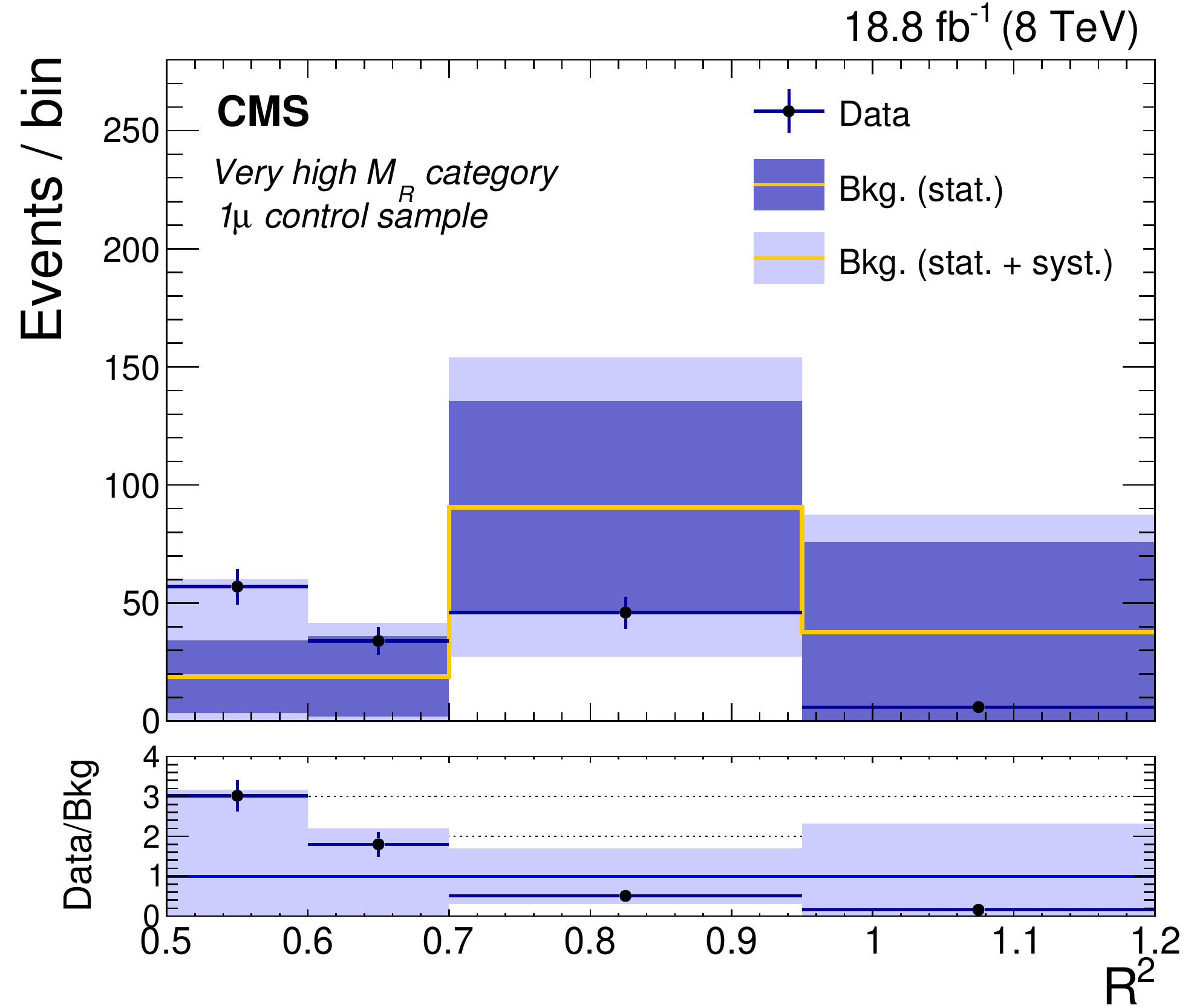}
 \caption{Comparison of observed yields in the 1$\mu$ control region and the
   data-driven background estimate derived from on the 2$\mu$ control region data in the four $\MR$ categories:
   VL (top left), L (top right), H (bottom left), and VH (bottom right). The bottom panel in each plot shows the
   ratio between the two distributions. The observed bin-by-bin
   deviation from unity is interpreted as an estimate of the systematic uncertainty
   associated to the background estimation methodology for the 0$\mu$ search region. The
   dark and light bands represent the statistical and the total
   uncertainties in the estimates, respectively. The horizontal bars indicate
the variable bin widths.\label{fig:1muCLOSURE}}
\end{figure}

Figure~\ref{fig:1muCLOSURE} shows the comparison of the
$\RR^2$ distributions between the observed yield and the
data-driven background estimate in the 1$\mu$ control
region. The observed bin-by-bin difference is propagated as
a systematic uncertainty in the data-driven background method,
and accounts for the statistical uncertainty in the
event yield in the 2$\mu$ control region data as well as
potential differences in the modeling of the recoil spectra
between $\PW$+jets and $\cPZ$+jets processes. Some bins exhibit
relatively large uncertainties primarily due to statistical fluctuations
in the 2$\mu$ control region from which the background is prediction estimated.
Though the uncertainties are rather large in fractional terms,
sensitivity to DM signal models is still obtained, because of the enhanced
signal to background ratio for the bins at large values of $\RR^2$.

\begin{figure}
  \centering
   \includegraphics[width=0.47\textwidth]{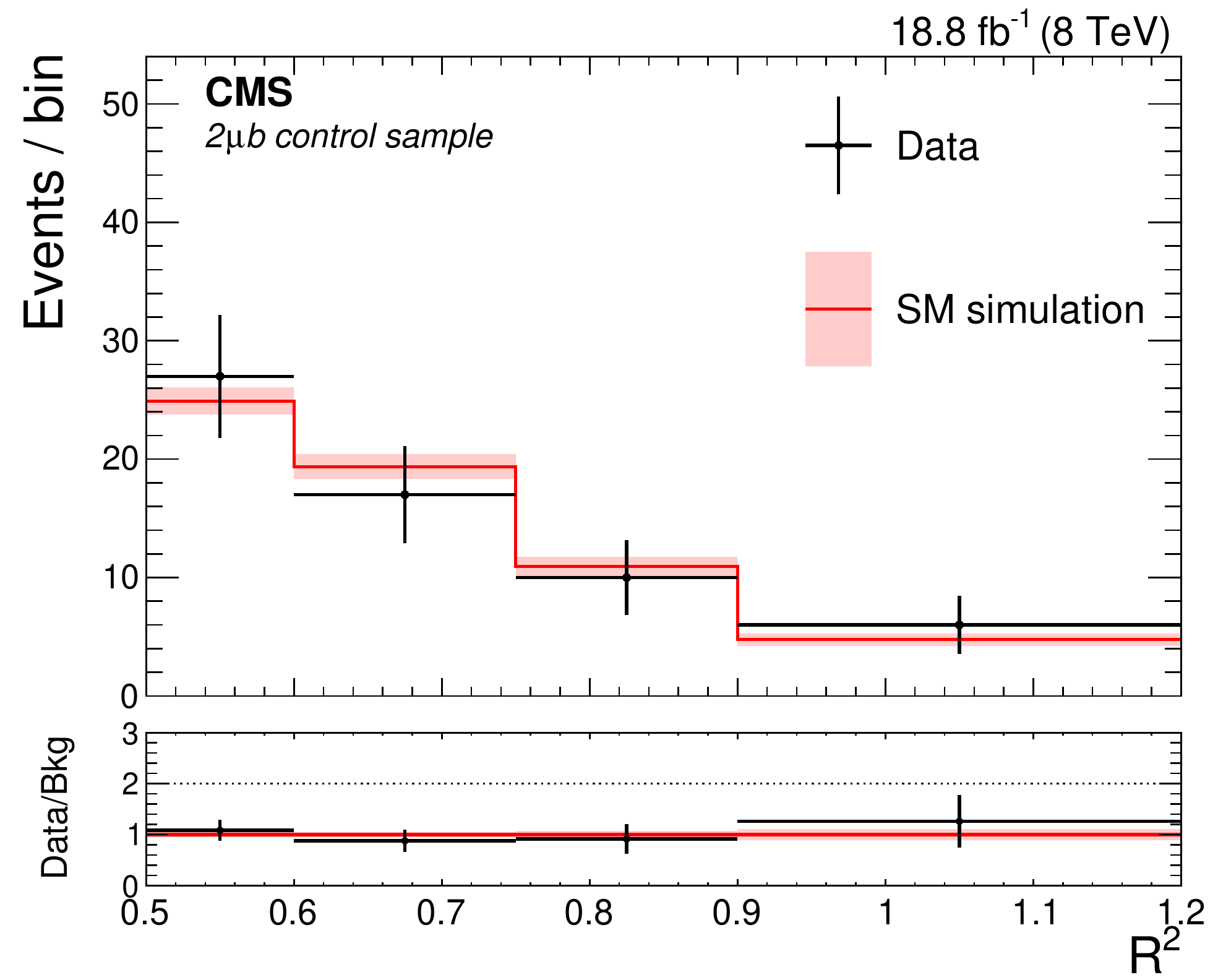}
   \caption{Comparison of the observed yield and the prediction from
     simulation as a function of $\RR^2$ in the 2$\mu$b control region.
     The uncertainties in the data and the simulated
     sample are represented by the vertical bars and the shaded bands,
     respectively. The horizontal bars indicate
the variable bin widths.\label{fig:2mub}}
\end{figure}

The $\ttbar$ background is estimated using an analogous data-driven method,
where we derive corrections to the Monte Carlo simulation prediction
scaled to the $\ttbar$ production cross-section computed to NNLO accuracy~\cite{WatNNLO,ZatNNLO,TTbaratNNLO}
using data in the 2$\mu$b control region for each bin in $\RR^2$.
The correction is then applied to the simulation prediction for
the $\ttbar$ background contribution to the zero b-tag search region.
This correction factor reflects potential mismodeling of the recoil
spectrum predicted by the Monte Carlo simulation.
The contribution of each background process to
the 2$\mu$b sample, predicted from simulated samples, is given in
Table~\ref{tab:2mub}. The fraction of \ttbar events in the 2$\mu$b
control sample is ${\approx}95\%$.
\begin{table}
\centering
\topcaption{\label{tab:2mub} Observed yield and predicted
    background from simulated samples in the 2$\mu$b control region.
    The quoted uncertainty in the prediction only reflects the size of the simulated sample.
    The contribution of each individual background process is also shown,
    as estimated from simulated samples.}
\begin{tabular}{*{7}{c}}
  \hline
  Sample  &  $\cPZ(\PGn\PAGn)$+jets  &  $\PW(\ell \nu)$+jets &
  $\cPZ(\ell \ell)$+jets  &  $\ttbar$  & MC predicted &  Observed \mT\mB\\
  \hline
  2$\mu$b  &   $<$0.1  & $0.1\pm0.1$ & $2.2\pm0.3$ & $58\pm2$ & $60\pm2$ & 60 \mT\mB\\
  \hline
\end{tabular}
\end{table}
Figure~\ref{fig:2mub} shows the comparison of the observed
yield and the prediction from simulation, as a function of
$\RR^2$. We observe no significant deviations between the observed
data and the simulation prediction. The uncertainty derived from the
data-to-simulation correction factor is propagated to the systematic uncertainty
of the $\ttbar$ prediction in the zero b-tag search region.

The result of the background estimation in the zero b-tag search region is given in
Table~\ref{tab:bkg0mu}, where it is compared to the observed yields in
data. The uncertainty in the background estimates takes into account
both the statistical and systematic components.

\begin{figure}
 \centering
   \includegraphics[width=0.48\textwidth]{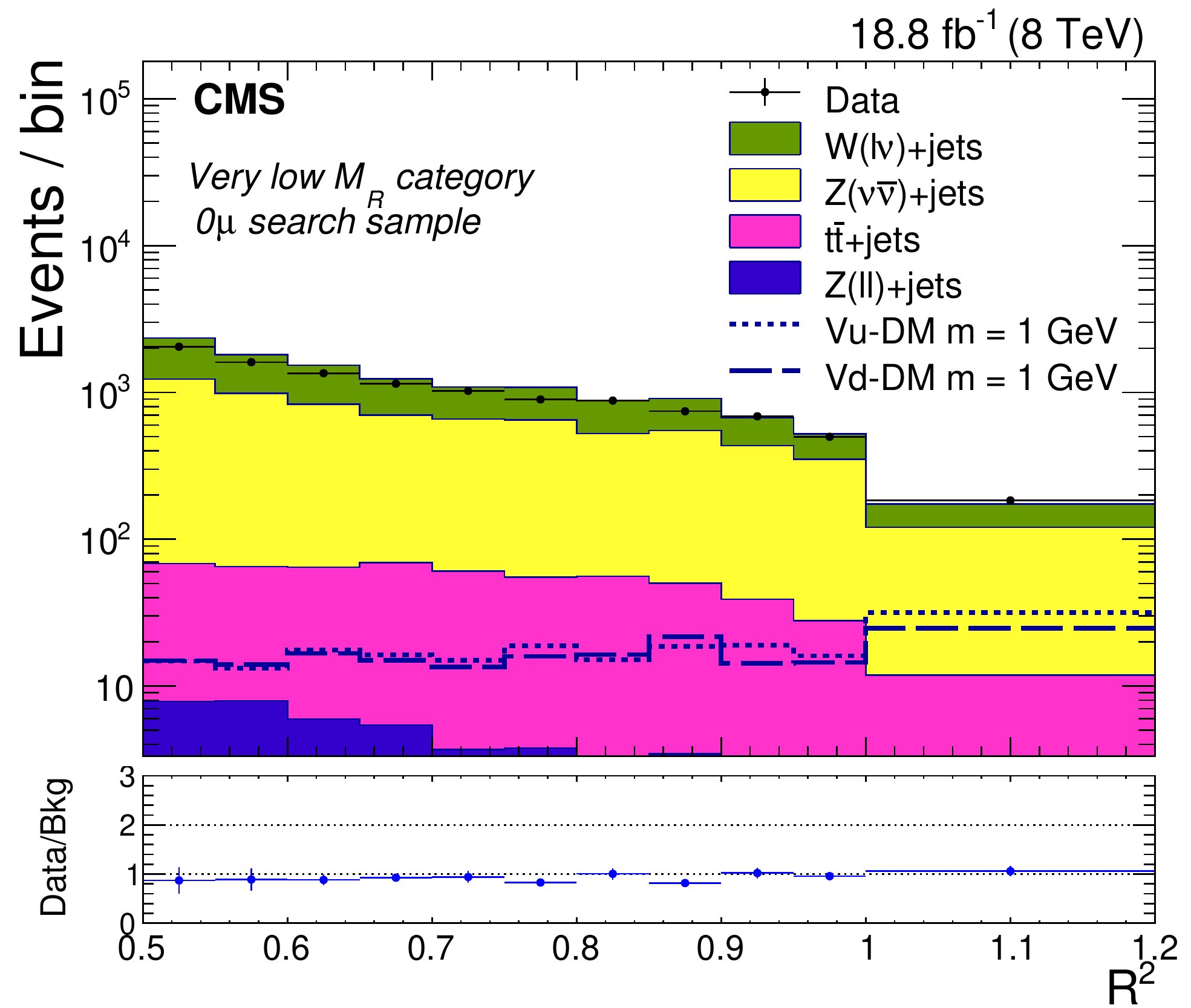}
   \includegraphics[width=0.48\textwidth]{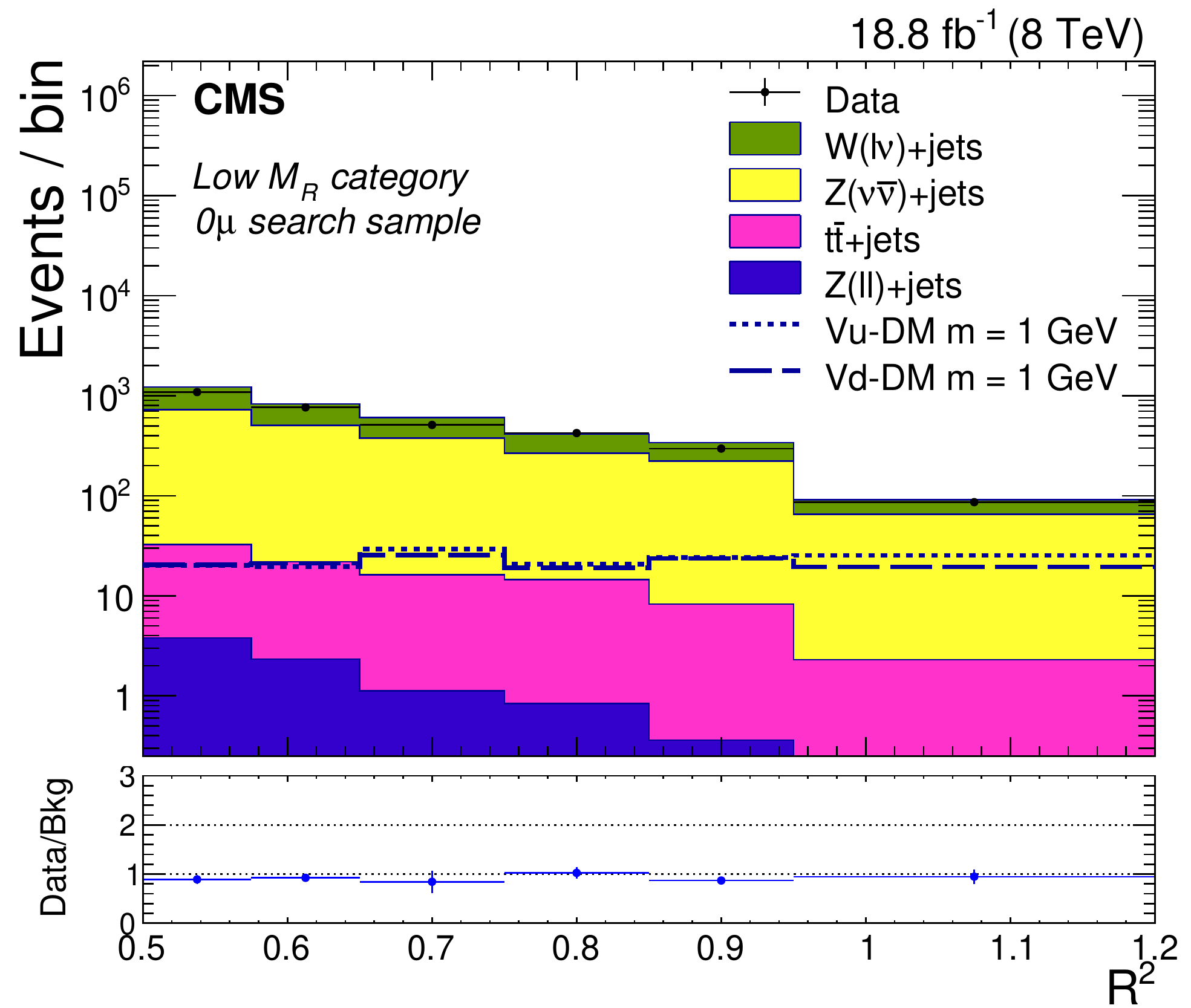}
   \includegraphics[width=0.48\textwidth]{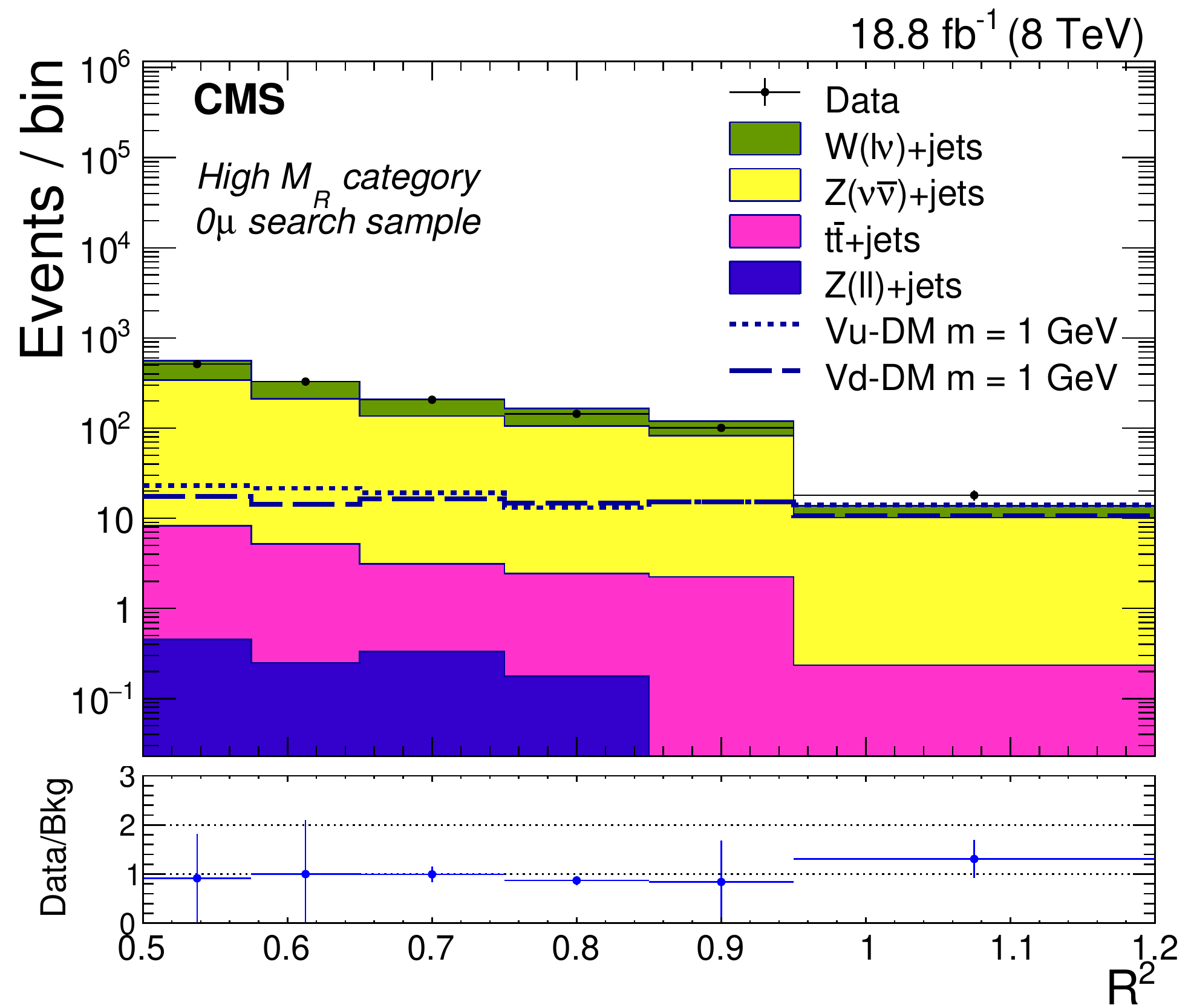}
   \includegraphics[width=0.48\textwidth]{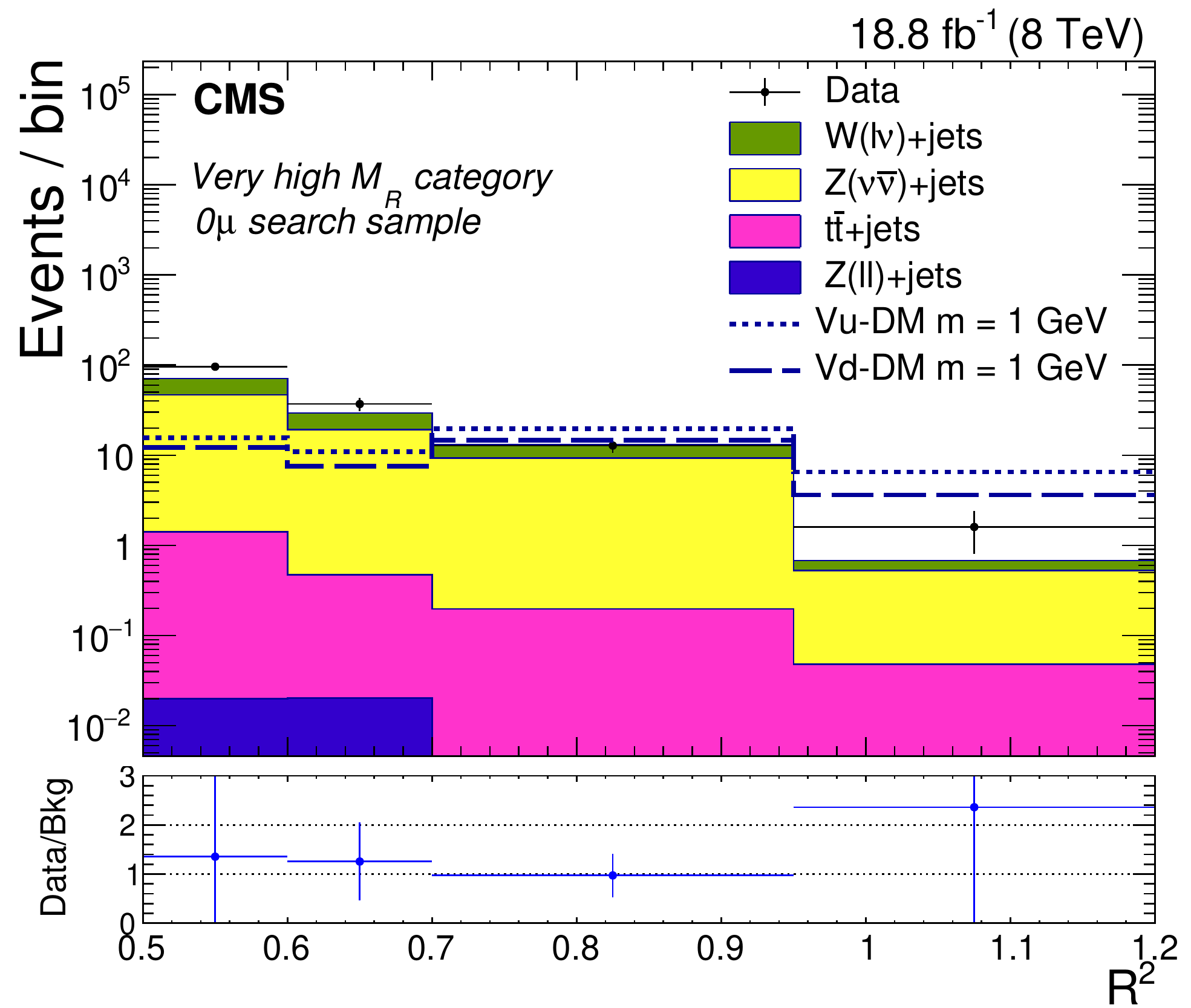}
 \caption{Comparison of the observed yield in the zero b-tag control region and the
   background estimates in the four $\MR$ categories:
   VL (top left), L (top right), H (bottom left), and VH (bottom right). The
   contribution of individual background processes is shown by the
   filled histograms. The bottom panels show the ratio
   between the observed yields and the total background estimate. The systematic
   uncertainty in the ratio includes the systematic uncertainty in the background estimate.
   For reference, the distributions from two
   benchmark signal models are also shown, corresponding to the pair
   production of DM particles of mass 1\GeV in the EFT approach with
   vector coupling to u or d quarks. The horizontal bars indicate
the variable bin widths.\label{fig:0muSignalBkg1GeV} }
\end{figure}

\begin{table}
\centering
\topcaption{\label{tab:bkg0mu}  Comparison of the observed yields for
  for the zero b-tag search region in each $\MR$ category and the
   corresponding background estimates. The uncertainty in
    the background estimate takes into account both the statistical
    and systematic components. The contribution of each individual background process is also shown, as
estimated from simulated samples, as well as the total MC predicted yield.}
\resizebox{\textwidth}{!}{
 \begin{tabular}{c*{6}{x}r}
  \hline
\multicolumn{1}{c}{$\MR$ category} &  \multicolumn{1}{c}{$\cPZ(\PGn\PAGn)$+jets}  &  \multicolumn{1}{c}{$\PW(\ell \nu)$+jets}  &  \multicolumn{1}{c}{$\cPZ(\ell \ell)$+jets}  &  \multicolumn{1}{c}{$\ttbar$}  & \multicolumn{1}{c}{MC predicted}&  \multicolumn{1}{c}{Estimated}  &  \multicolumn{1}{c}{Observed} \mT\mB\\
  \hline
  VL  &  6231,37  & 4820,33 & 49,2 & 555,7 & 11655,50 &12770,900 & 11623 \mT\\
  L  &  2416,19 & 1513,16 & 11,1 & 104,3 & 4044,25 & 4170,270 & 3785 \\
  H  &  1127,7 & 625,9 & 2.9,0.3 & 24,1  &  1779,12 & 1650,690 & 1559 \\
  VH  &  229,2 & 103,3 & 0.2,0.1 & 3.1,0.5 &
   335,3 & 240,160 & 261\mB\\
  \hline
\end{tabular}
}
\end{table}
The comparison of the data-driven background estimates and the observations for
each $\MR$ category is shown in Fig.~\ref{fig:0muSignalBkg1GeV}, as a function of $\RR^2$.
The expected event distribution is shown for two signal
benchmark models, corresponding to the pair production of DM particles
of mass 1\GeV in the effective field theory (EFT) approach with vector coupling to u or d quarks.
Details on the signal benchmark models are given in Section~\ref{sec:EFT0mu}.

\subsection{Background estimation for the \texorpdfstring{0$\mu$b and 0$\mu$bb}{0 mu b and 0 mu bb} samples}

A similar data-driven technique is used to determine the expected background for the
one and two b-tag search regions. The background from $\ttbar$ events for each $\RR^2$ bin in the
one b-tag search region, $n(\ttbar)_{i}^{0\mu\PQb}$, is computed as:
\begin{equation}
  n(\ttbar)_{i}^{0\mu\PQb} =  \bigl(n(\ttbar)_{i}^{2\mu\PQb} - N_{i}^{\cPZ(\ell
    \ell)+\text{jets},2\mu\PQb} - N_{i}^{\PW(\ell\nu)+\text{jets},2\mu\PQb}\bigr)
\frac{N(\ttbar)_{i}^{0\mu\PQb}}{N(\ttbar)_{i}^{2\mu\PQb}}\label{eq:bBkgPred}
\end{equation}
where $n(\ttbar)_{i}^{2\mu\PQb}$ is the observed yield in the $i$th $\RR^2$
bin in the 2$\mu$b control region, while $N(\ttbar)_{i}^{0\mu\PQb}$ and
$N(\ttbar)_{i}^{2\mu\PQb}$ are the $\ttbar$ yields in the $i$th $\RR^2$ bin
predicted by the simulation for the one b-tag search region and the 2$\mu$b control region
respectively. Similarly, the \ttbar background in the two b-tag search region
is derived from Eq.~(\ref{eq:bBkgPred}), replacing $N(\ttbar)_{i}^{0\mu\PQb}$
with $N(\ttbar)_{i}^{0\mu\mathrm{bb}}$, the \ttbar
background yield in the $i$th bin of the two b-tag search region predicted
by the simulation. The data yield in the 2$\mu$b control region is
corrected to account for the small contamination from $\cPZ$+jets and
$\PW$+jets, predicted with the simulated yields $N_{i}^{\cPZ(\ell\ell)+\text{jets},2\mu\PQb}$ and
$N_{i}^{\PW(\ell\nu)+\text{jets},2\mu\PQb}$, respectively.

\begin{figure}
\centering
   \includegraphics[width=0.48\textwidth]{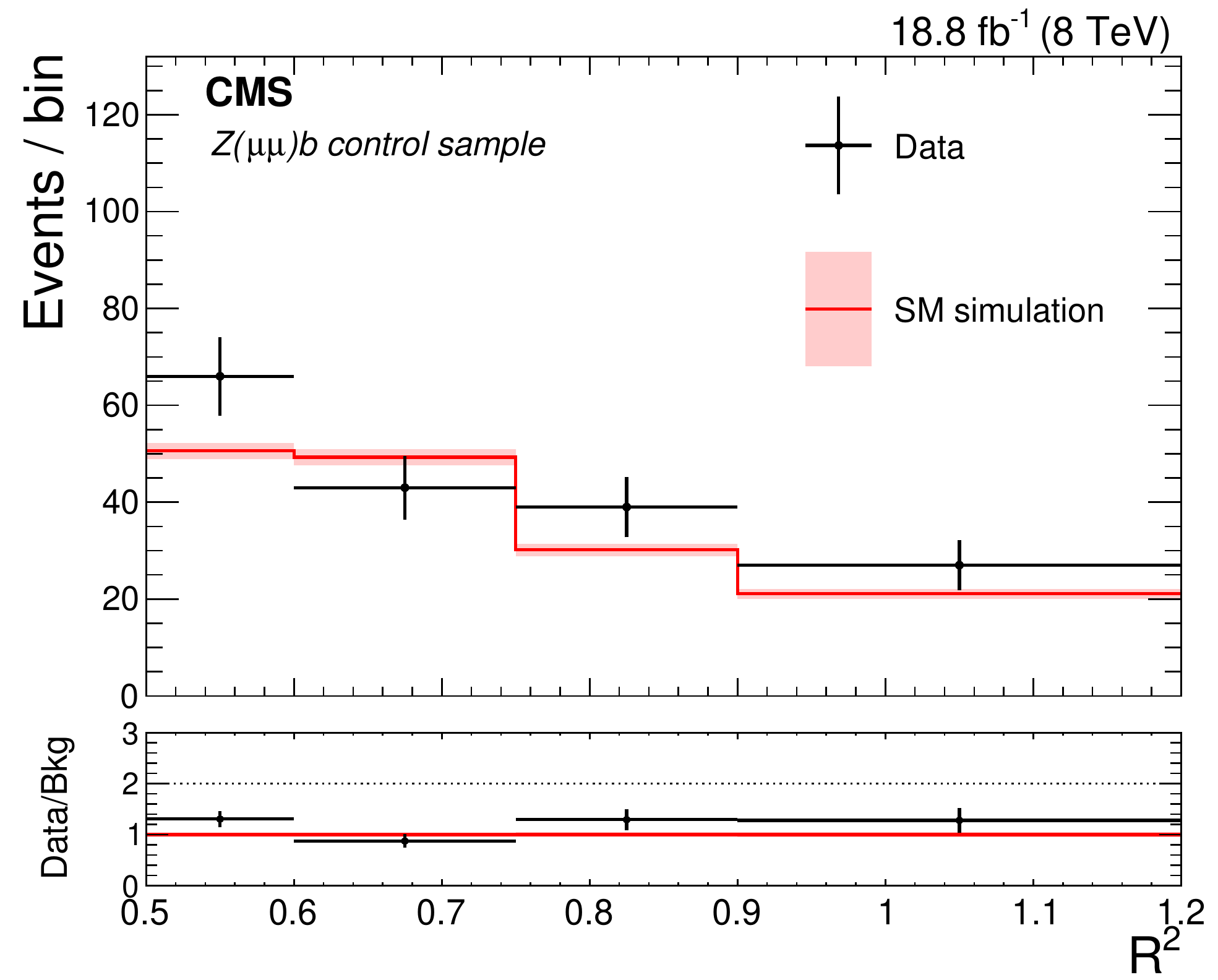}
   \includegraphics[width=0.48\textwidth]{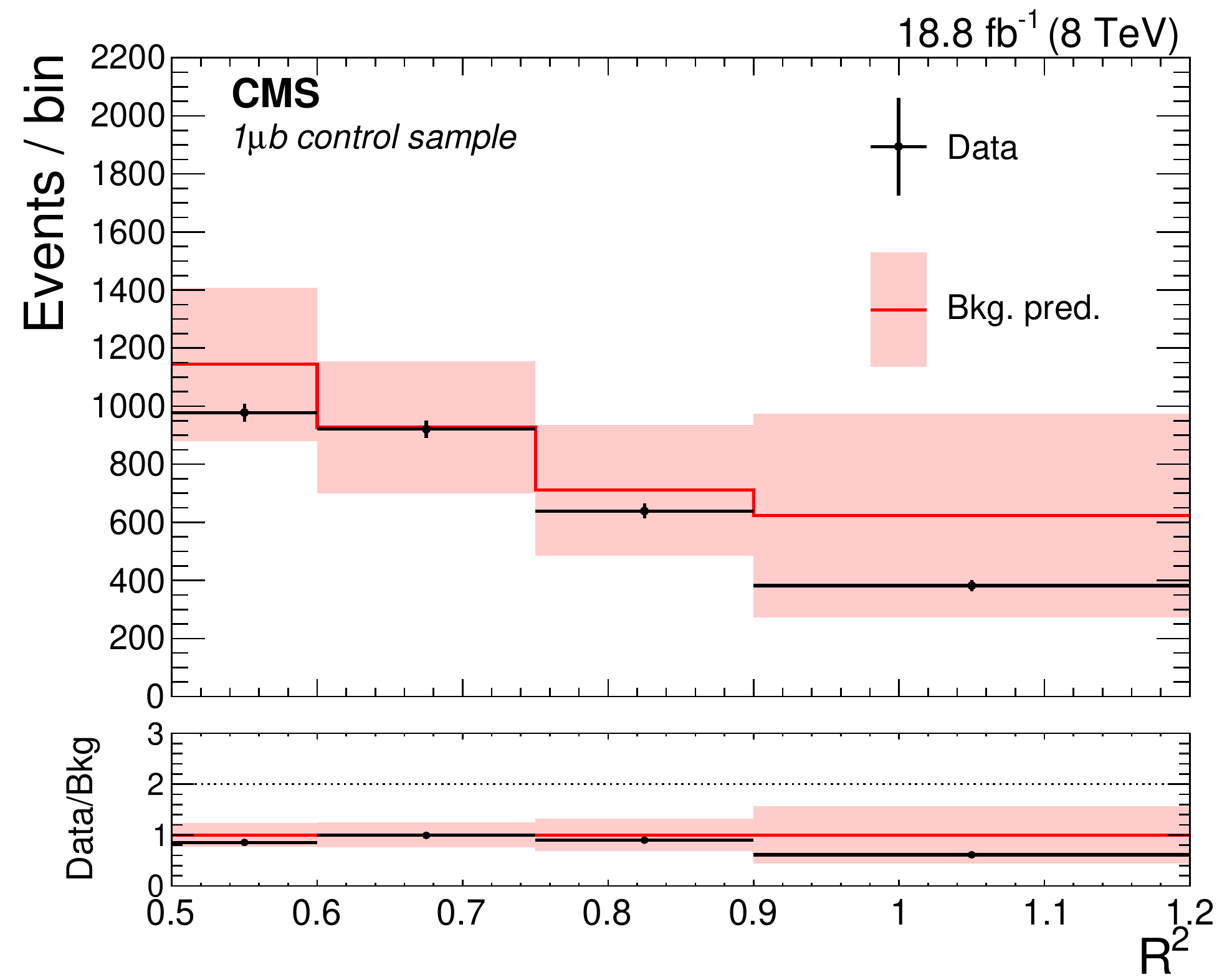}
 \caption{Comparison of the observed yield and the
   prediction from simulation in the $\cPZ(\mu\mu)$b control sample (left)
   and of the observed yield in the $1\mu$b control sample and
   the background estimates from the 2$\mu$b and $\cPZ(\mu\mu)$b
   control samples (right), shown as a function of $\RR^2$. The bottom
   panel of each figure shows the ratio between the data and the
   estimates. The shaded bands represent the statistical uncertainty
   in the left plot, and the total uncertainty in the right plot. The horizontal bars indicate
the variable bin widths.\label{fig:Zmumub}}
\end{figure}

The background contribution from $\PW(\ell \nu)$+jets and $\cPZ(\PGn\PAGn)$+jets events is
predicted using the $\cPZ(\mu\mu)$b control region, and summarized in Table~\ref{tab:WITHB}.
The $\cPZ$+jets purity of this control region is ${\approx}$89\%.
The observed yield in the $\cPZ(\mu\mu)$b control region is shown in the left plot of Fig.~\ref{fig:Zmumub},
as a function of $\RR^2$, along with the Monte Carlo simulation prediction. The uncertainty on the
simulation prediction accounts only for the statistical uncertainty of the simulated sample.
This contribution, scaled by the ratio of the predicted V+jets background in the search regions
to that in the control region, obtained from simulation, provides an estimate for each $\RR^2$ bin.

\begin{table}
\centering
\topcaption{\label{tab:WITHB}
    Comparison of the observed yields in the $\cPZ(\mu\mu)$b and
    $1\mu$b samples, the corresponding predictions from background
    simulation, and (for $1\mu$b only) the cross-check background
    estimate. The contribution of each individual background process is also shown, as
estimated from simulated samples.}
\resizebox{\textwidth}{!}{
\begin{tabular}{*{7}{c}r}
  \hline
  Sample  &  $\cPZ(\PGn\PAGn)$+jets  &  $\PW(\ell \nu)$+jets  &
  $\cPZ(\ell \ell)$+jets  &  $\ttbar$   &  MC predicted  & Estimated &
  \multicolumn{1}{c}{Observed} \mT\mB\\
  \hline
   $\cPZ(\mu\mu)$b  &  $<$0.1  &  $<$0.1  & $134\pm3$ & $17\pm1$ &
   $151\pm3$ &\NA  &  175 \mT\mB\\
   $1\mu$b &  $0.2\pm0.1$ & $279\pm7$ & $11\pm1$ & 3038 $\pm$
   17 & $3328\pm18$ & $3410\pm540$ & 2920\mT\mB\\
  \hline
\end{tabular}
}
\end{table}

We perform a cross-check of the method on the 1$\mu$b control region
by predicting the background from the 2$\mu$b control region data.
The data and prediction are compared on the right of Fig.~\ref{fig:Zmumub},
where we observe reasonable agreement. The difference between the
prediction and the observed data in this cross-check region is propagated as a
systematic uncertainty of the method.

\begin{figure}
   \includegraphics[width=0.48\textwidth]{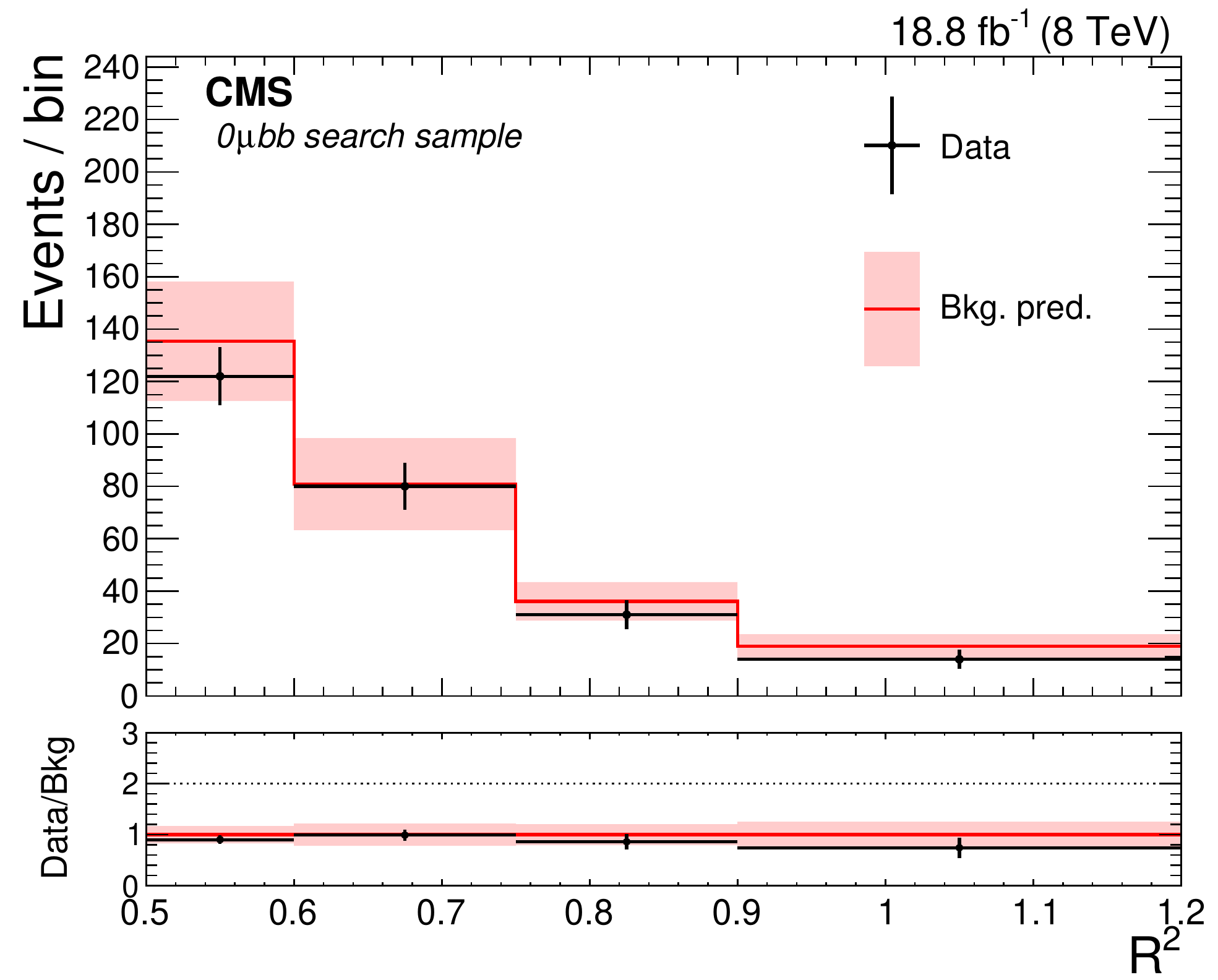}
   \includegraphics[width=0.48\textwidth]{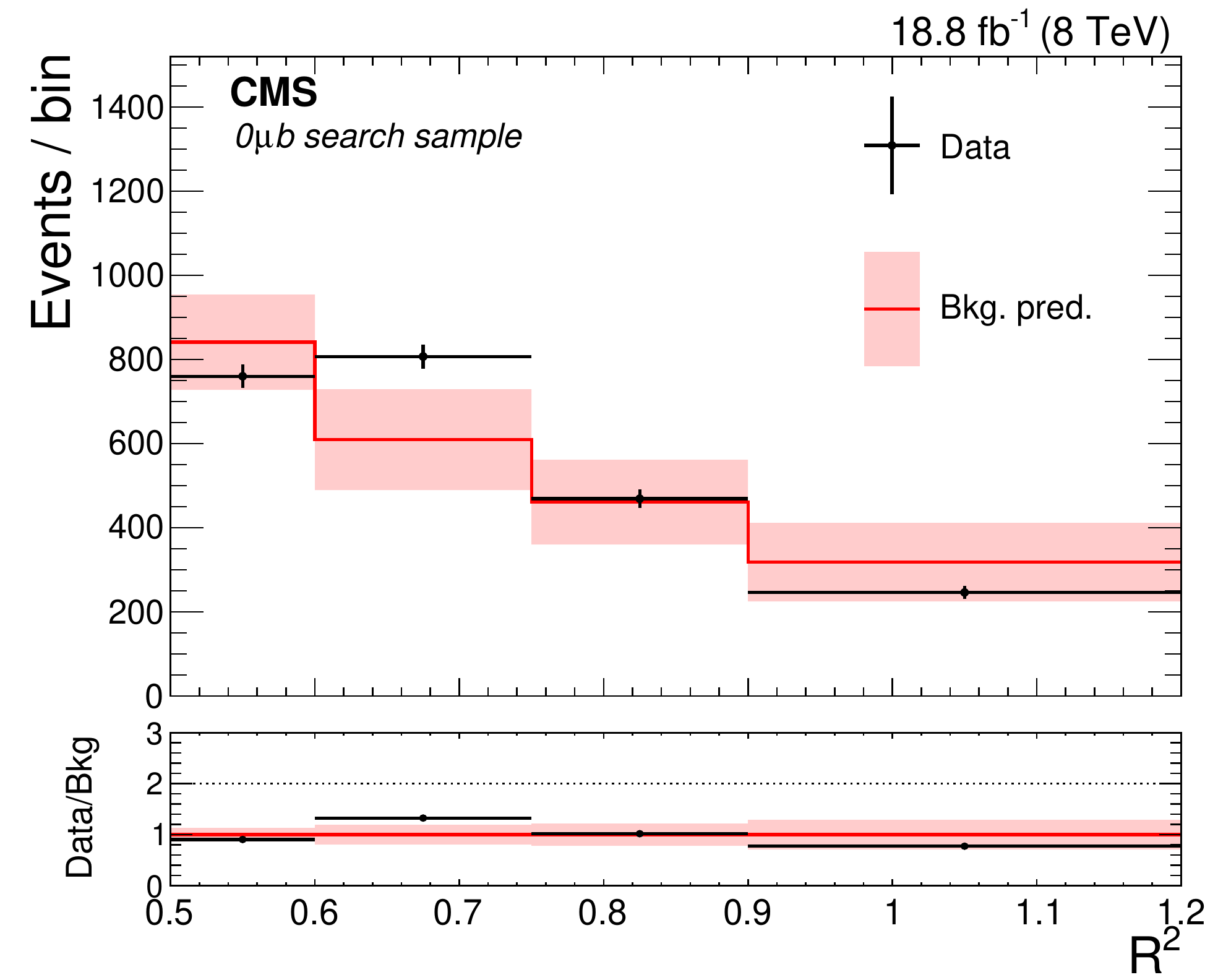}
 \caption{Comparison of observed event yields and background
   estimates as a function of $\RR^2$, for the
   one (left) and two (right) b-tag search regions.
   The shaded bands represent the total uncertainty in the estimate. The horizontal bars indicate
the variable bin widths.\label{fig:0muttbar}}
\end{figure}

The estimated background in the one and two b-tag search regions
is given in Table~\ref{tab:bkg0muWITHB} and shown in Fig.~\ref{fig:0muttbar}, where it is compared to
the observed yields in data. The uncertainty in the
estimates take into account both the statistical and systematic
components.
\begin{table}
\centering
  \topcaption{ Comparison of the observed yield for
    events in the one and two b-tag search regions and the corresponding background
    estimates. The uncertainty in the estimates takes into account
    both the statistical and systematic components. The contribution of each individual background process is also shown, as
estimated from simulated samples, as well as the total MC predicted yield.
\label{tab:bkg0muWITHB}}
\resizebox{\textwidth}{!}{
\begin{tabular}{*{7}{c}r}
  \hline
 Sample  &  $\cPZ(\PGn\PAGn)$+jets  &  $\PW(\ell \nu)$+jets  &
  $\cPZ(\ell \ell)$+jets  &  $\ttbar$  & MC predicted& Estimated &  \multicolumn{1}{c}{Observed} \mT\mB\\
\hline
  $0\mu$bb  &  $44\pm3$  &  $14\pm2$  &  $0.2\pm0.1$  &  204
  $\pm$ 4  & $262\pm5$ &  $271\pm37$  &  247 \mT\mB\\
  $0\mu$b  &  $417\pm8$  &  $216\pm7$  &  $2.4\pm0.4$  &  $1480\pm12$  & $2115\pm16$ &  $2230\pm280$  &  2282 \mT\mB\\
\hline
\end{tabular}
}
\end{table}

\section{Systematic uncertainties}\label{sec:sys}

For each $\RR^2$ bin in each $\MR$ category, the
difference between the observed and estimated yields in the
crosscheck analysis (see Section~\ref{sec:bkg}) is taken as the estimate of the uncertainty associated 
with the method, and covers the differences in the modeling of the recoil spectra 
between $\PW$+jets and $\cPZ$+jets processes as well as the cross section uncertainties. 
These uncertainties are found to be typically ${\approx}$20--40\%, depending on the considered bin in the
($\MR$, $\RR^2$) plane, and are the dominant systematic uncertainties for the analysis.
As discussed in Section~\ref{sec:bkgzmu}, a few bins at smaller values of $\RR^2$ exhibit 
larger systematic uncertainties, primarily due to statistical fluctuations in the
control region. However the impact on the sensitivity to the dark matter models considered
is small as the signal to background ratio is significantly better in other bins at larger 
values of $\RR^2$.

For the 0$\mu$ analysis, differences between the kinematic properties of $\PW$+jets and
$\cPZ$+jets events are additional sources of systematic uncertainty. These
differences arise from the choice of the PDF set, jet energy scale
corrections, b tagging efficiency
corrections, and trigger efficiency. These effects largely cancel when taking the ratio of the two processes, and the resulting
uncertainty is found to be smaller than one fifth
of the total uncertainty.  The quoted uncertainty is an upper
estimate of the total systematic uncertainty.

For the 0$\mu$b and 0$\mu$bb samples, both the signal and control samples are dominated by \ttbar events. The cancellation of the
systematic uncertainties is even stronger in
this case, since it does not involve different processes, and different
PDFs. The remaining uncertainty is
dominated by the contribution arising from the small size of the control
sample.

Systematic uncertainties in the signal simulation
originate from the choice of the PDF set,
the jet energy scale correction, the modeling of the initial-state radiation in the event
generator, and the uncertainty in the integrated luminosity. The
luminosity uncertainty changes the signal normalization while the other
uncertainties also modify the signal shape.
These effects are taken into account by propagating these
uncertainties into the $\MR$ category and the $\RR^2$ bin. These uncertainties are
considered to be fully correlated across $\MR$ categories and
$\RR^2$ bins. Typical values for the individual contributions
are given in Table~\ref{tab:sygSys}. The total uncertainty in the
signal yield is obtained by propagating the individual effects into
the $\MR$ and $\RR^2$ variables and comparing the bin-by-bin variations with respect to the central value of the prediction
based on simulation. In the particular case of the uncertainties due
to the choice of the PDF set we have followed the PDF4LHC~\cite{Bourilkov:2006cj,Alekhin:2011sk,Botje:2011sn} prescription, using the CTEQ-6.6\cite{Nadolsky:2008zw} and MRST-2006-NNLO~\cite{Martin:2007bv} PDF sets.

\begin{table}
 \centering
 \topcaption{\label{tab:sygSys} Systematic uncertainties associated with
  the description of the DM signal. The values indicated represent
  the typical size. The dependence of these systematic uncertainties on
  the $\RR^2$ and $\MR$ values is taken into
  account in the determination of the results.}
 \begin{tabular}{ll}
  \hline
  \multicolumn{1}{c}{Effect}  &  \multicolumn{1}{c}{Uncertainty}\\
  \hline
  Jet energy scale  &  3--6\%\\
  Luminosity  &  2.6\%\\
  Parton distribution functions  &  3--6\%\\
  Initial-state radiation  &  8--15\%\\
  \hline
\end{tabular}
\end{table}

\section{Results and interpretation}\label{sec:interpretation}

In Figs.~\ref{fig:0muSignalBkg1GeV}~and~\ref{fig:0muttbar} the
estimated backgrounds are compared to the observed yield in each
$\MR$ region, for events without and with
b-tagged jets, respectively. The background estimates agree with the observed
yields, within the uncertainties. This result is interpreted in terms of exclusion limits
for several models of DM production.

\subsection{Limits on dark matter production from the \texorpdfstring{0$\mu$}{0 mu} sample}\label{0muResults}
\label{sec:EFT0mu}
The result is interpreted in the context of a low-energy effective
field theory, in which
the production of DM particles is mediated by six or seven dimension
operators~\cite{maverickDM,TevatronDMFrontier}. This choice allows
the results be compared with those of previous
analyses~\cite{Aad:2011xw,Chatrchyan:2012me},
and shows that a similar sensitivity is achieved.

Operators of dimension six and seven are generated assuming the
existence of a heavy particle, mediating the interaction between the
DM and SM fields. To describe DM production as a local interaction,
the propagator of the heavy mediator is expanded through an operator
product expansion. The nature of the mediator
determines the nature of the effective interaction. Two benchmark
scenarios are considered in this study, axial-vector (AV), and vector
(V) interactions~\cite{PhysRevD.85.056011}, described by the
following operators:
\begin{equation}
\label{eq:OvOva}
\hat{\mathcal{O}}_{\mathrm{AV}}
=
\frac{1}{\Lambda^{2}}\left(\bar{\chi}\gamma^{\mu}\gamma_{5}\chi\right)
\left(\bar{q}\gamma_{\mu}\gamma_{5}q\right) \hspace{0.05in};\qquad
\hat{\mathcal{O}}_{\mathrm{V}} =
\frac{1}{\Lambda^{2}}\left(\bar{\chi}\gamma^{\mu}\chi\right)
\left(\bar{q}\gamma_{\mu}q\right).
\end{equation}
Here  $\gamma_{\mu}$ and $\gamma_{5}$ are the Dirac matrices, $\chi$
is the DM field, and $q$ is an SM quark field. The DM particle is assumed to be a Dirac
fermion where both operators will contribute in the low-energy theory, while in the case of a Majorana DM particle the vector coupling
$\hat{\mathcal{O}}_{V}$ will vanish in the low-energy theory.
Below the cutoff energy
scale $\Lambda$, DM production is described as a contact interaction
between two quarks and two DM particles. In the case of $s$-channel
production through a heavy mediator, the energy scale $\Lambda$ is
identified with $M/g_\text{eff}$, where $M$ is the mediator mass and
$g_\text{eff} = \sqrt{g_{q} g_{\chi}}$ is an effective
coupling, determined by the coupling of the mediator to quark and DM
fields, $g_q$ and $g_\chi$, respectively.

The results in Tables~\ref{tab:LOOKUP_VL}-\ref{tab:LOOKUP_VH} in the
Appendix are used to obtain an upper limit at 90\% confidence level
(CL) on the DM production cross section, $\sigma^{i}_{\mathrm{UL}}$ (where
the superscript denotes the coupling to an up or down quark). The
limits are obtained using the
LHC CL$_\mathrm{s}$ procedure~\cite{LHC_CLS,CMS-NOTE-2011-005} and a global likelihood determined
by combining the likelihoods of the different search categories. Each
 systematic uncertainty (see Section~\ref{sec:sys}) is incorporated in the likelihood with a dedicated nuisance parameter, whose value is not known a priori but rather must be estimated from the data.

\begin{figure}
\centering
\includegraphics[width=0.48\textwidth, angle=0.]{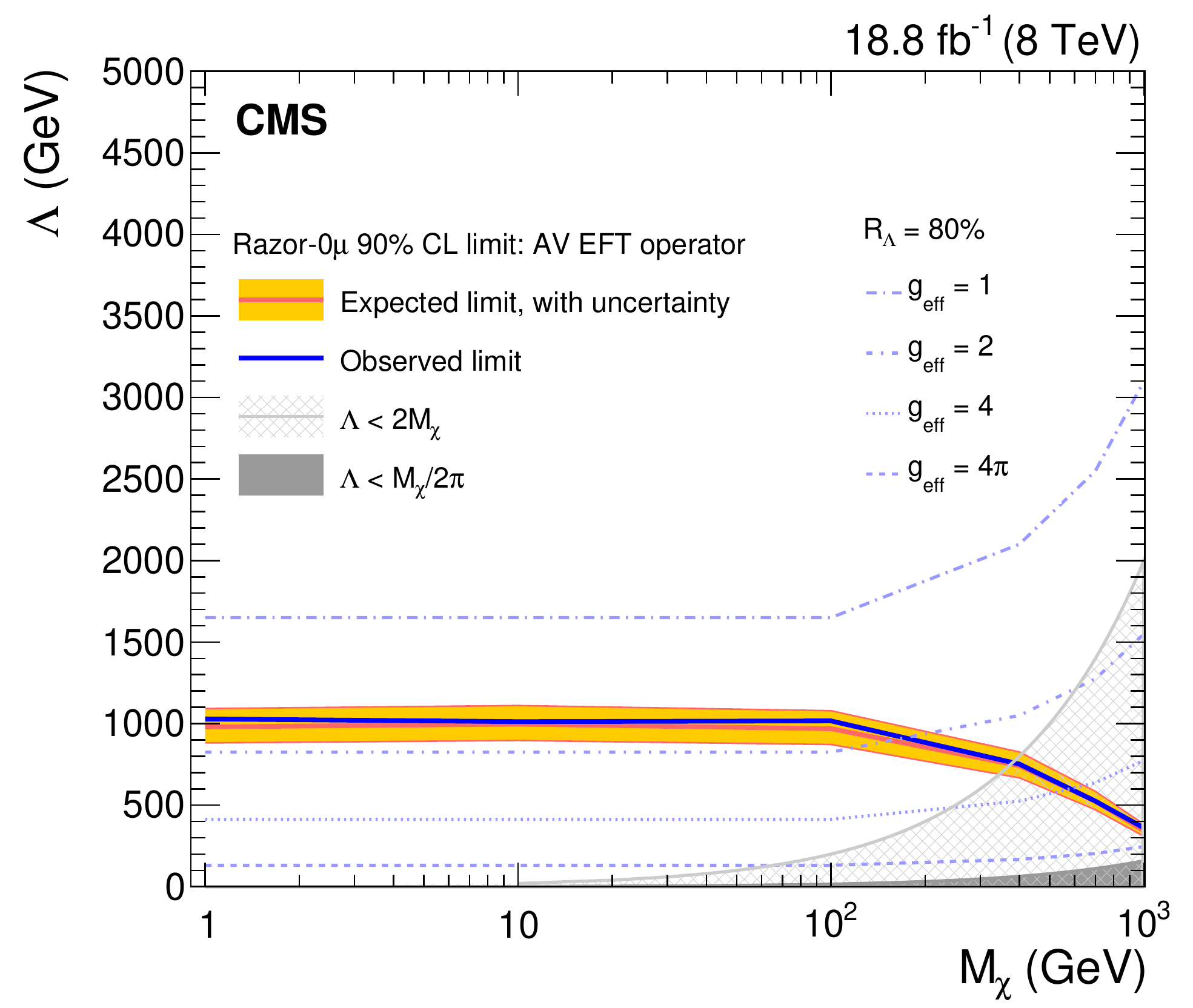}
\includegraphics[width=0.48\textwidth]{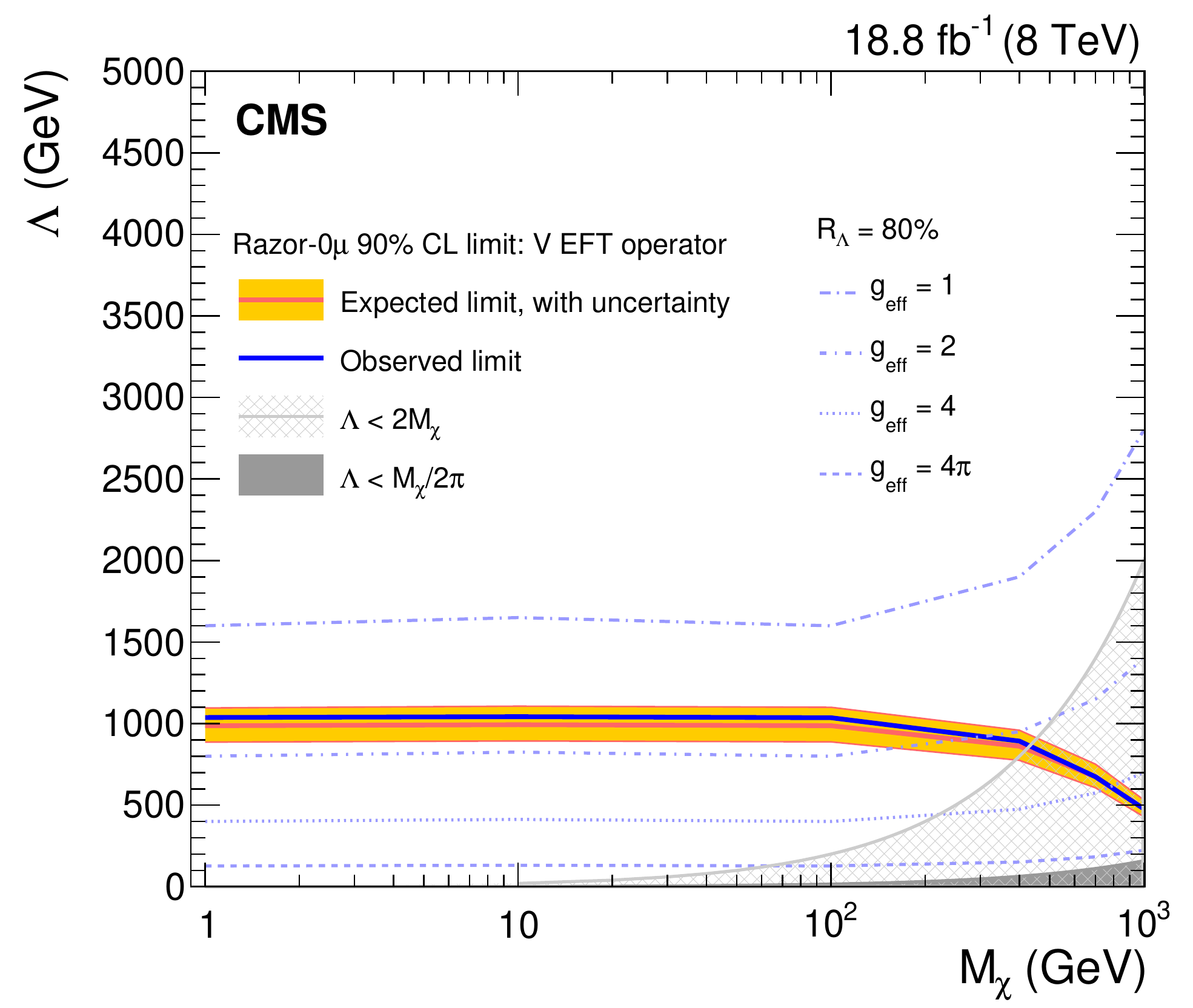}
\caption{Lower limit at 90\% CL on the cutoff scale $\Lambda$ as a
  function of the DM mass $M_\chi$ in the case of
  axial-vector (left) and vector (right) currents. The validity of the EFT
  is quantified by $R_\Lambda = 80\%$ contours, corresponding to
  different values of the effective coupling
  $g_\text{eff}$. For completeness, regions forbidden by the
  EFT validity condition $\Lambda > 2M_\chi/g_\text{eff}$  are shown for two choices of the
effective coupling:  $g_\text{eff} = 1$ (light gray) and $g_\text{eff}= 4\pi$ (dark gray).\label{fig:LambdaLimit}}
\end{figure}

Subsequently, the cross section ($\sigma^{i}_{\mathrm{UL}}$) limit is translated into a lower limit $\Lambda_{\mathrm{LL}}$ on
the cutoff scale, through the relation:
\begin{equation}
\Lambda_\mathrm{LL} = \Lambda_\text{GEN} \left(\frac{\sigma_\text{GEN}}{\sigma_\mathrm{UL}}\right)^\frac{1}{4}.
\end{equation}
Here $\Lambda_\text{GEN}$ and $\sigma_\text{GEN}$ are the cutoff
energy scale and cross section of the simulated sample, respectively.
The derived values of $\Lambda_\mathrm{LL}$ as a function of the DM mass,
shown in Fig.~\ref{fig:LambdaLimit}, are very similar to those derived
for the CMS monojet search~\cite{monojet8TeV}. The
exclusion limits on $\Lambda$ weaken at large DM masses since the
cross section for DM production is reduced. The analysis has been
repeated removing the events also selected by the monojet search.  The
reduction in background yields due to this additional requirement
compensates for the reduction in signal efficiency, resulting in a
negligible difference in the exclusion limit on $\Lambda$.

The EFT framework provides a benchmark scenario to compare
the sensitivity of this analysis with that of previous searches for
similar signatures. However, the
validity of an EFT approach is limited at the LHC because a fraction
of events under study are generated at a $\sqrt{\hat s}$ comparable to
the cutoff scale $\Lambda$~\cite{Goodman:2010ku,TevatronDMFrontier,Friedland:2011za,Buchmueller:2013dya}. For theories to be perturbative, $g_\text{eff}$ is
typically required to be smaller than $4\pi$, and this condition is
unlikely to be satisfied for the entire region of phase space probed by
the collider searches. In addition, the range
of values for the couplings being probed within the EFT may be unrealistically large. Following
the study presented in Refs.~\cite{Riotto1,Riotto2,Riotto3}, we
quantify this effect through two EFT validity measures. The first is a
minimal kinematic constraint on $\Lambda$ obtained by requiring
$Q_\text{tr} < g_\text{eff}\Lambda$ and $Q_\text{tr} >
2M_{\chi}$, where $Q_\text{tr}$ is the momentum transferred
from the mediator to the DM particle pair, which yields $\Lambda > 2M_{\chi}/g_\text{eff}$ . The second is more stringent and uses the quantity:
\begin{equation}
R_\Lambda = \frac{\int \rd\RR^2 \int \rd \MR
  \displaystyle{\left.\frac{\rd^2\sigma}{\rd\RR^2  \rd \MR}\right\vert_{Q_\text{tr}<g_\text{eff}\Lambda} }}
{\int \rd\RR^2 \int \rd \MR \displaystyle{\frac{\rd^2\sigma}{\rd\RR^2  \rd\MR}}}.
\end{equation}

Values of $R_\Lambda$ close to unity indicate a regime in which the
assumptions of the EFT approximation hold, while a deviation from unity quantifies
the fraction of events for which the EFT approximation is still valid. We
consider the case of $s$-channel production, and we compute
$R_\Lambda$ as a function of the effective coupling $g_\text{eff}$
in the range $0 < g_\text{eff} \leq 4\pi$.  The contours
corresponding to $R_\Lambda = 80\%$ for different values of
$g_\text{eff}$ are shown in Fig.~\ref{fig:LambdaLimit}. For values
of $g_\text{eff} \gtrapprox 2$, the limit set by the analysis lies
above the $R_\Lambda = 80\%$ contour.

\begin{figure}
\centering
\includegraphics[width=0.48\textwidth]{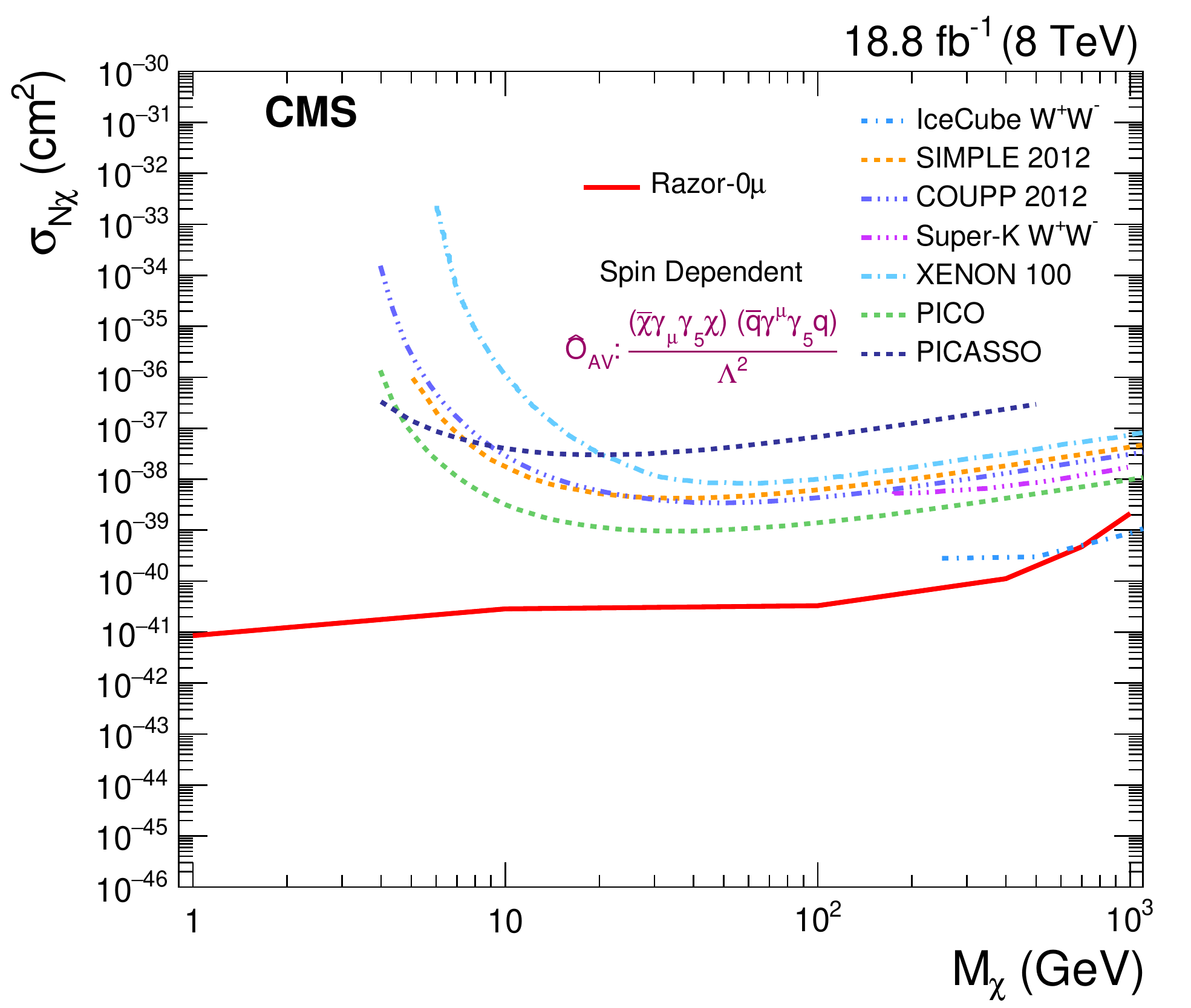}
\includegraphics[width=0.48\textwidth]{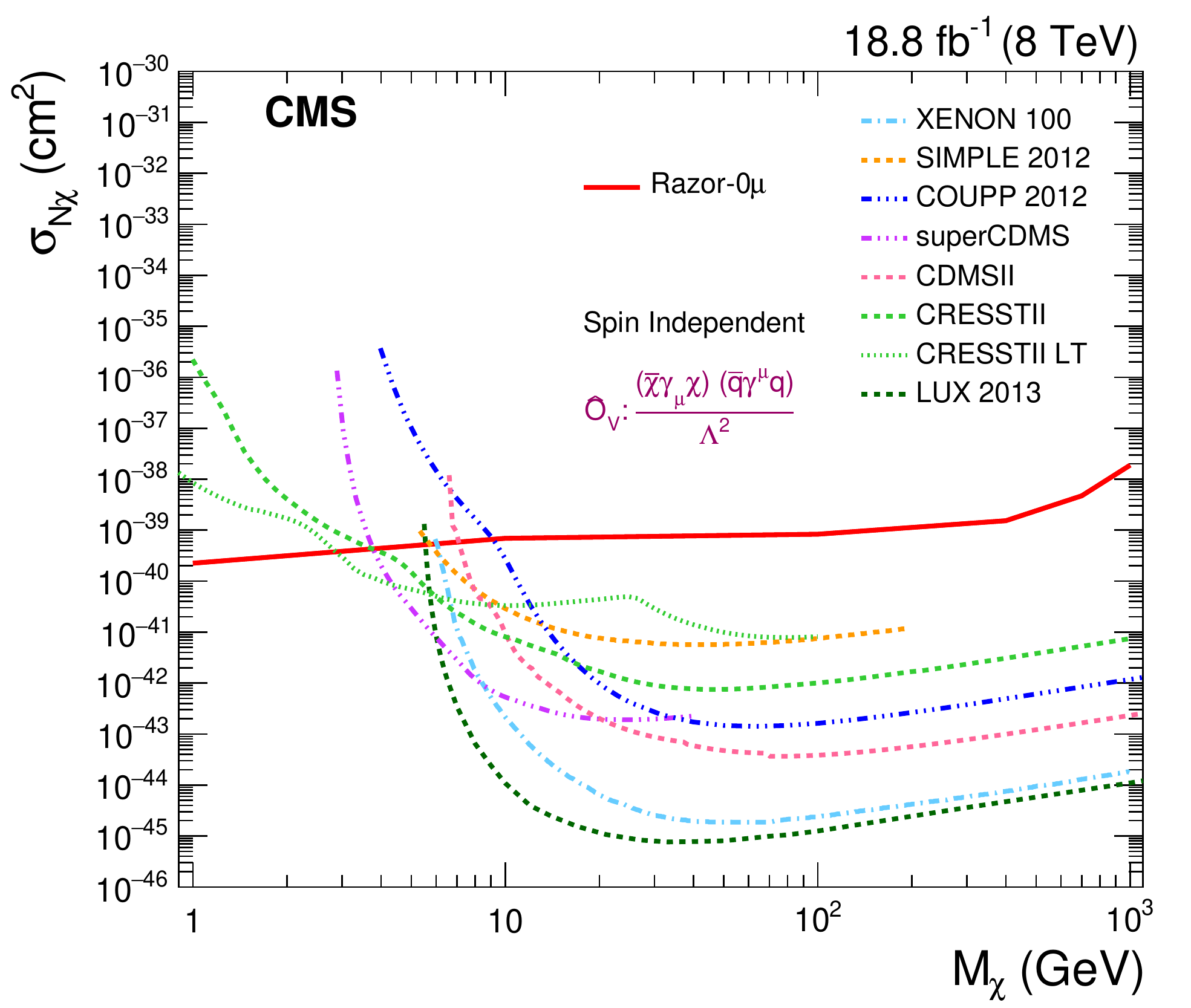}
\caption{Upper limit at 90\% CL on the DM-nucleon scattering cross
  section $\sigma_{N\chi}$ as a function of the DM mass
  $M_\chi$ in the case of spin-dependent axial-vector (left)
  and spin-independent vector (right) currents. A selection of
  representative direct detection experimental bounds are also shown.\label{fig:DMNxsec}}
\end{figure}

The exclusion limits on $\Lambda$ for the axial-vector and vector operators are transformed into upper limits on
the spin-dependent ($\sigma^\mathrm{SD}_{N\chi}$)~\cite{Super-Kamiokande,
  IceCube, COUPP, SIMPLE, Amole:2015lsj, Archambault:2012pm, Aprile:2013doa} and spin-independent
($\sigma^\mathrm{SI}_{N\chi}$)~\cite{SIMPLE, COUPP, CDMS-II, SuperCDMS, XENON100, LUX, Angloher:2014myn, Angloher:2015ewa}
DM-nucleon scattering cross section, respectively; using the following
expressions~\cite{PhysRevD.85.056011}:
\begin{align}
\sigma_{N\chi}^\mathrm{SD}  & =  0.33 \frac{\mu^{2}}{\pi\Lambda_\mathrm{LL}^{4}}, \\
\sigma_{N\chi}^\mathrm{SI}  & =  9 \frac{\mu^2}{\pi\Lambda_\mathrm{LL}^{4}},\\
\intertext{where}
\mu &= \frac{M_{\chi}M_{\Pp}}{M_{\chi}+M_{\Pp}},
\end{align}
with $M_{\Pp}$ and $M_\chi$ indicating the proton and
DM masses, respectively. The numerical values of the derived limits
are given in Tables~\ref{tab:AVLimit}~and~\ref{tab:VLimit}.  The
bound on $\sigma_{N\chi}$ as a function of $M_\chi$ is shown
in Fig.~\ref{fig:DMNxsec} for spin-dependent and spin-independent
DM-nucleon scattering. A summary of the observed limits for the
axial-vector and vector operators can be found in Tables~\ref{tab:AVLimit} and~\ref{tab:VLimit}
respectively. It is observed that the spin-independent bounds obtained by direct detection experiments are more stringent than those obtained by
the present result for masses above $\simeq 5$\GeV. Such an effect is expected since
the spin-independent DM-nucleus cross section is enhanced by the coherent
scattering of DM off nucleons in the case of spin-independent
operators. We note that the present result is more sensitive for small
DM mass because the recoil energy in direct detection experiments is lower in
this region and therefore more difficult to detect. In the case of spin-dependent
DM-nucleus scattering, the present results are more stringent that those
obtained by direct detection experiments because the DM-nucleus cross section
does not benefit from the coherent enhancement. A summary of the observed limits for the
axial-vector and vector operators can be found in
Tables~\ref{tab:AVLimit} and~\ref{tab:VLimit} respectively.

\begin{figure}
\centering
\includegraphics[width=0.48\textwidth]{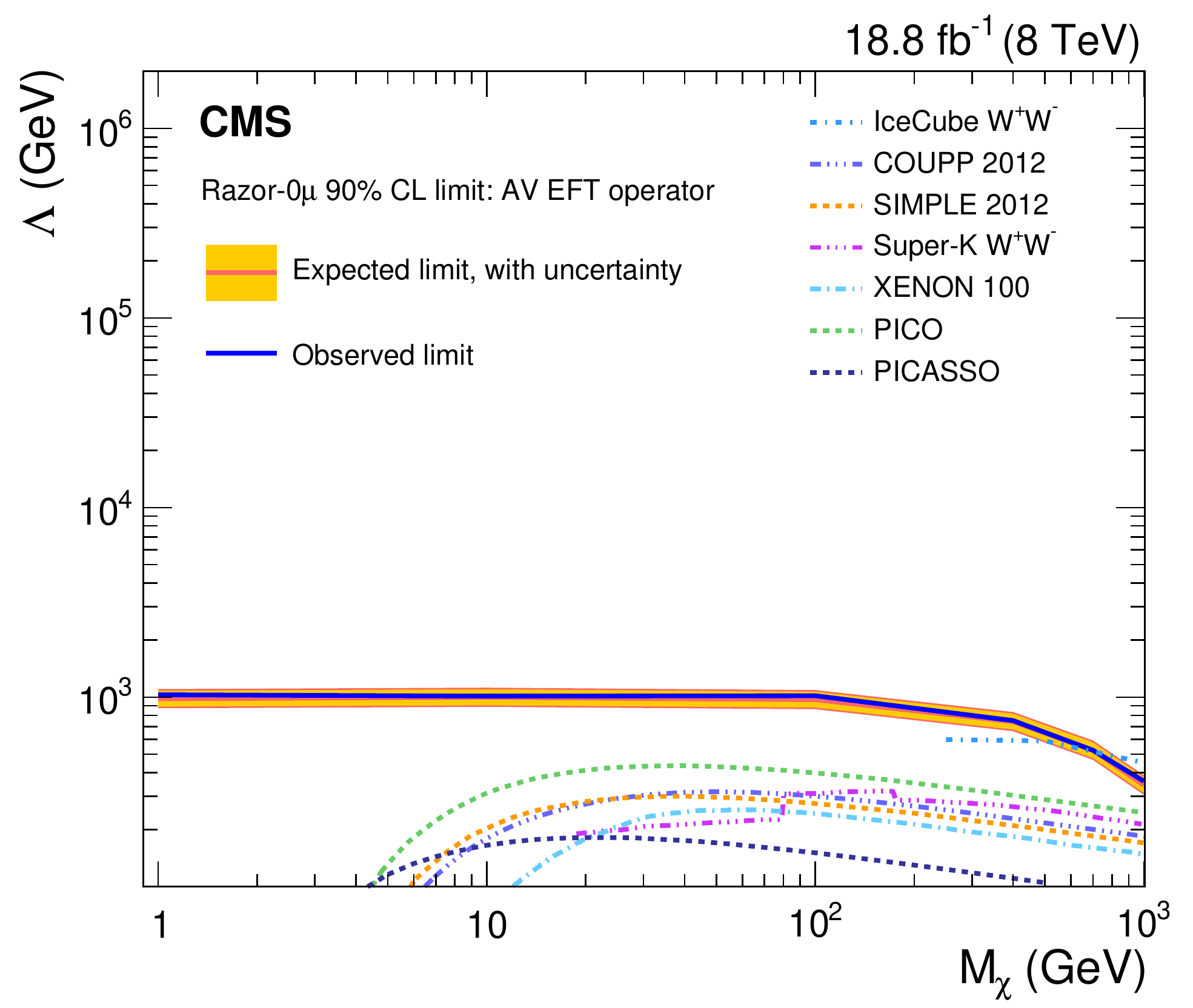}
\includegraphics[width=0.48\textwidth]{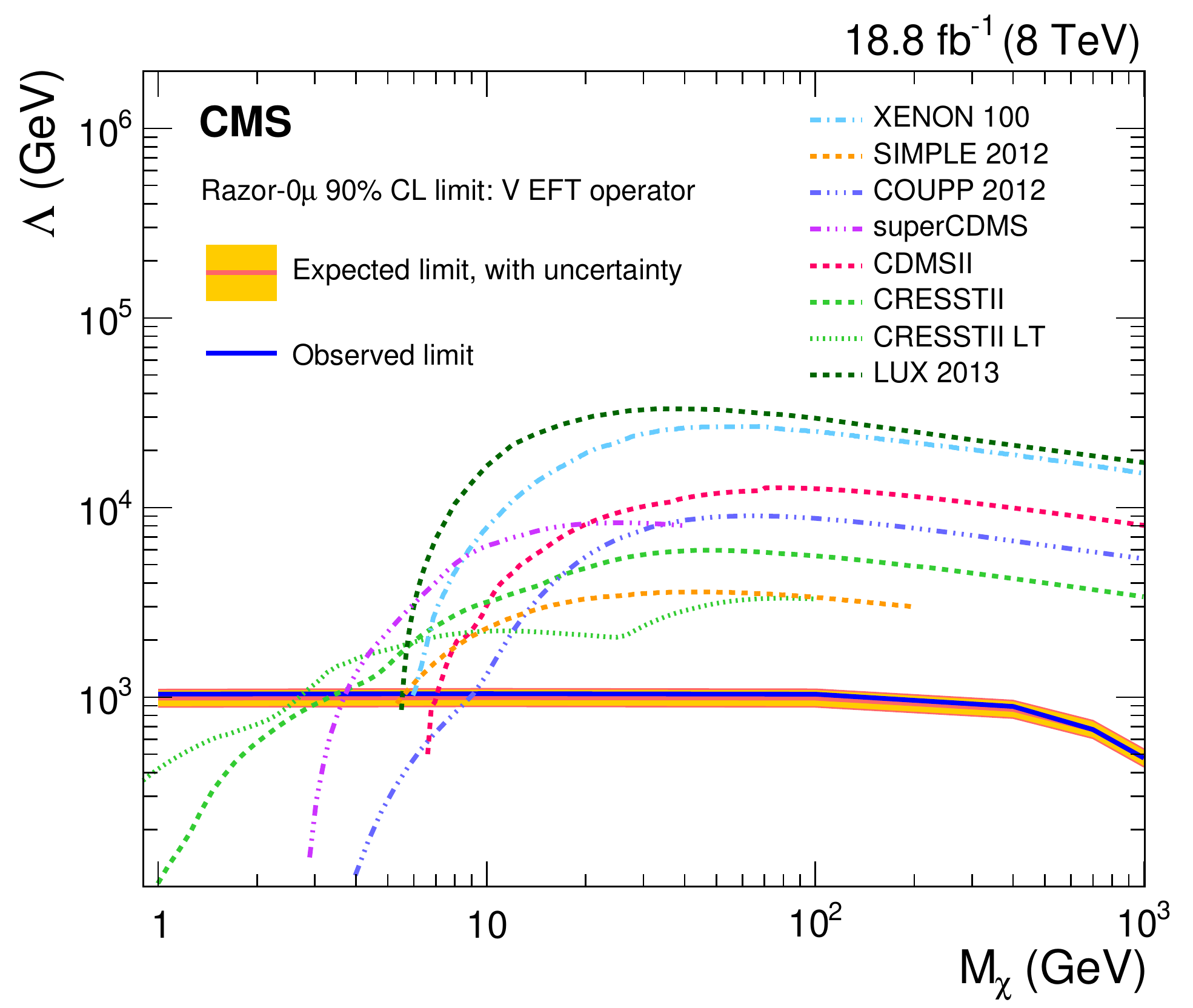}
\caption{Lower limit at 90\% CL on the cutoff scale $\Lambda$ as a
  function of the DM mass $M_\chi$ in the case of
  axial-vector (left) and vector (right) currents. A selection of direct detection
  experimental bounds are also shown.\label{fig:LambdaComplete}}
\end{figure}

In order to compare our results with those from direct detection experiments, the experimental bounds in~\cite{SIMPLE, COUPP,CDMS-II,
  SuperCDMS, XENON100, LUX,Super-Kamiokande, IceCube, COUPP, SIMPLE}
are translated into bounds on $\Lambda$. This comparison is shown in
Fig.~\ref{fig:LambdaComplete}. This translation is well
defined since the momentum transfer in most direct detection
experiments is low compared to the values of $\Lambda$ being probed,
and thus the EFT approximations in question are mostly valid.
\begin{table}
\centering
\topcaption{\label{tab:AVLimit}
The 90\% CL limits on DM production in the case of axial-vector
couplings. Here, $\sigma^{u}_\mathrm{UL}$ and $\sigma^{d}_\mathrm{UL}$ are
the observed upper limits on the production cross
section for u and d quarks, respectively; $\Lambda_\mathrm{LL}$
is the observed cutoff energy scale lower limit; and $\sigma_{N\chi}$
is the observed DM-nucleon scattering cross section upper limit.}
\begin{tabular}{r*{4}{c}}
\hline
\multicolumn{1}{c}{$M_\chi$  (\GeVns{})} &  $\sigma^{u}_\mathrm{UL}$(pb)  &  $\sigma^{d}_\mathrm{UL}$(pb)
  &  $\Lambda_\mathrm{LL}$ (\GeVns{})  &  $\sigma_{N\chi}$  $(\text{cm}^{2})$ \mT\mB\\
\hline
1  & 0.39  &  0.45  &  1029 & $8.5\times 10^{-42}$\mT \\
10  &  0.43  &  0.45   & 1012 & $2.9\times 10^{-41}$\\
100  &  0.30  & 0.37  &  1017 & $3.3\times 10 ^{-41}$\\
400  & 0.25  &  0.26  &  752 & $1.1\times 10^{-40}$\\
700  &  0.21  &  0.26  &  524 & $4.7\times 10^{-40}$\\
1000  & 0.17  & 0.22  &  360 & $2.1\times 10^{-39}$\\
\hline
\end{tabular}
\end{table}
\begin{table}
\centering
\topcaption{\label{tab:VLimit}
The 90\% CL limits on DM production in the case of vector
couplings. Here, $\sigma^{u}_\mathrm{UL}$ and $\sigma^{d}_\mathrm{UL}$ are
the observed upper limits on the production cross
section for u and d quarks, respectively; $\Lambda_\mathrm{LL}$
is the observed cutoff energy scale lower limit; and $\sigma_{N\chi}$
is the observed DM-nucleon scattering cross section upper limit.}
\begin{tabular}{*{5}{c}}
\hline
$M_\chi$  (\GeVns{}) &  $\sigma^{u}_\mathrm{UL}$(pb)  &  $\sigma^{d}_\mathrm{UL}$(pb)
  &  $\Lambda_\mathrm{LL}$ (\GeVns{})  &  $\sigma_{N\chi}$  $(\text{cm}^{2})$\mT\mB\\
\hline
1  &  0.41  &   0.38  & 1038 & $2.3\times 10^{-40}$\mT \\
10  &  0.36  &  0.45  & 1043 & $6.9\times 10^{-40}$\\
100  &  0.33  &   0.44 & 1036 & $8.3\times 10 ^{-40}$\\
400  &  0.23  &  0.35  & 893 & $1.5\times 10^{-39}$\\
700  &  0.22  &  0.27  & 674 & $4.7\times 10^{-39}$\\
1000  &  0.22  &  0.27  & 477 & $1.8\times 10^{-38}$\\
\hline
\end{tabular}
\end{table}
\subsection{Limits on dark matter production from the \texorpdfstring{0$\mu$b and 0$\mu$bb}{0 mu b and 0 mu bb} samples}

The results from the 0$\mu$b and 0$\mu$bb samples are interpreted in
an EFT scenario, following a methodology similar to that of
Section~\ref{sec:EFT0mu}. In this case, a heavy scalar mediator
 is considered~\cite{Lin:2013sca}, generating an operator:
\begin{equation}
\label{eq:Os}
\hat{\mathcal{O}}_{S} = \frac{M_{q}}{\Lambda^{3}}\bar{\chi}\chi \bar{q}q.
\end{equation}

The dependence on the mass, induced by the scalar nature of the
mediator, implies a stronger coupling to third-generation quarks,
enhancing the sensitivity of the 0$\mu$b and 0$\mu$bb samples to this
scenario. Unlike the case of V and AV operators, the production cross
section for this process is proportional to $1/\Lambda^{6}$. The
value of $\Lambda_\mathrm{LL}$ is then derived as
\begin{equation}
\Lambda_\mathrm{LL} = \Lambda_\text{GEN} \left(\frac{\sigma_\text{GEN}}{\sigma_\mathrm{UL}}\right)^{\frac{1}{6}}.
\end{equation}

\begin{figure}
\centering
  \includegraphics[width=0.48\textwidth]{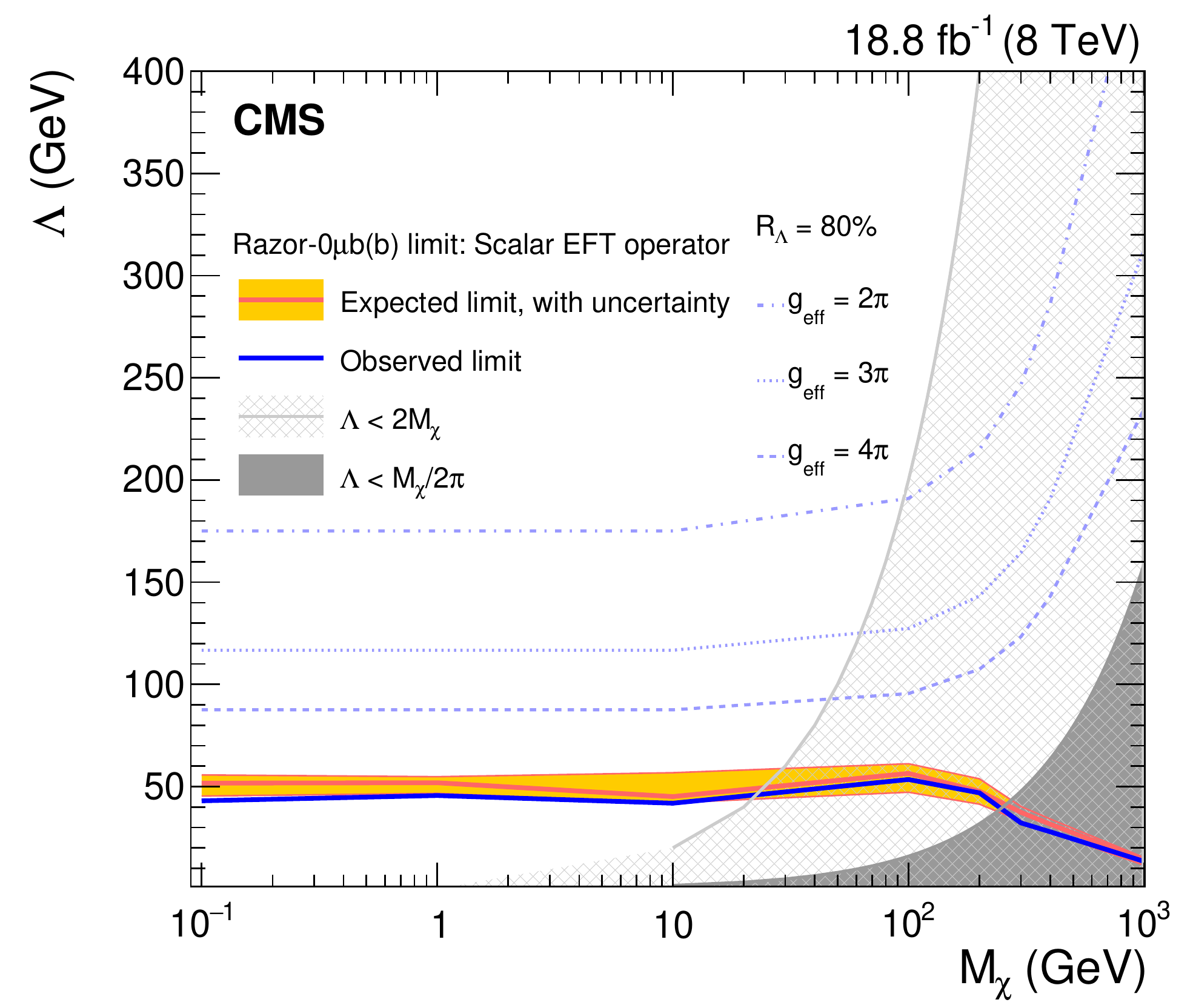}
  \includegraphics[width=0.48\textwidth]{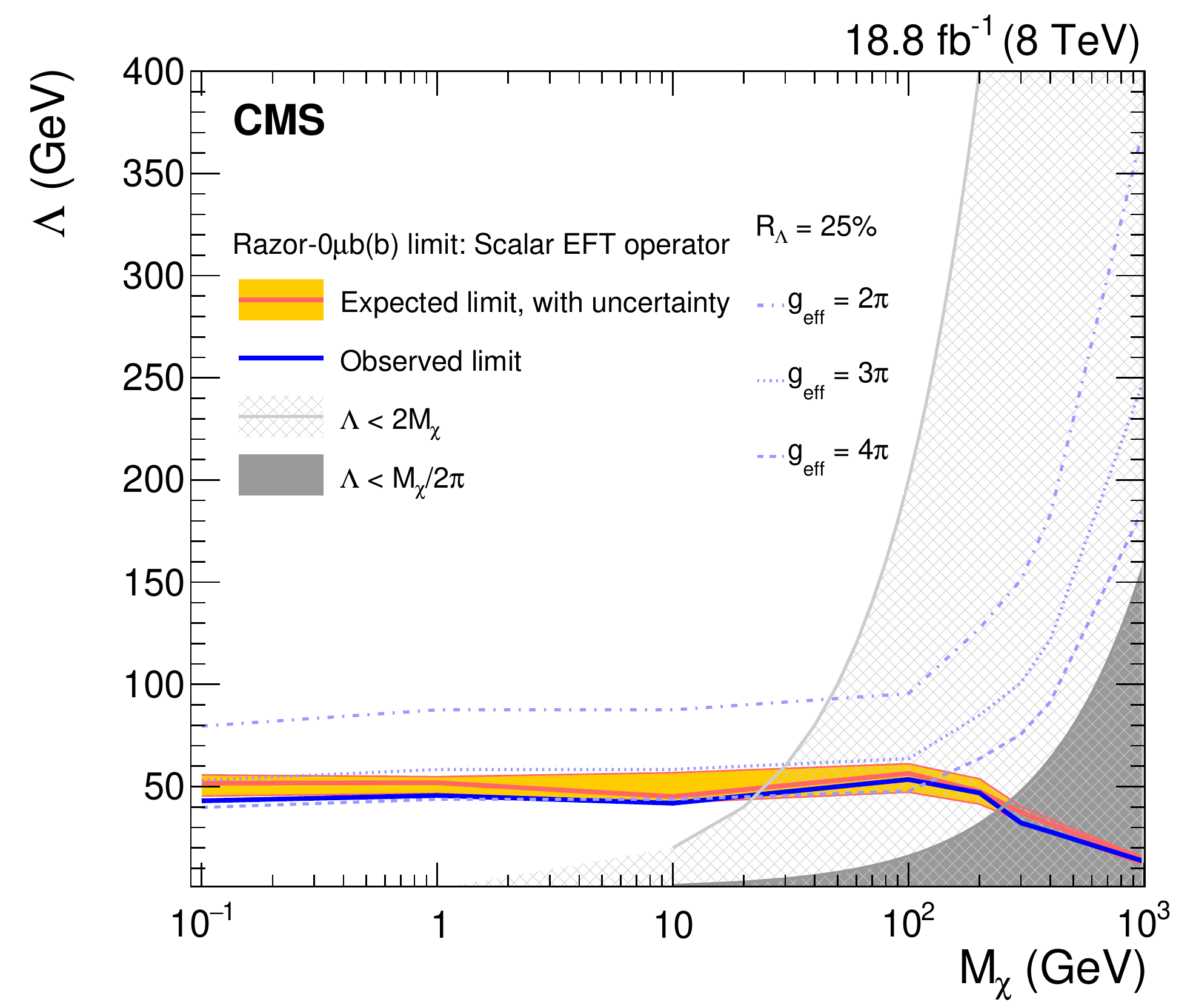}
\caption{Lower limit at 90\% CL on the cutoff scale $\Lambda$ for the
  scalar operator $\hat{\mathcal{O}}_{S}$ as a function of the DM mass
  $M_{\chi}$. The validity of the EFT is quantified by
  $R_\Lambda = 80\%$ (left) and $R_\Lambda = 25\%$ (right) contours,
  corresponding to different values of the effective coupling
  $g_\text{eff}$. For completeness, regions forbidden by the
  EFT validity condition $\Lambda > 2M_\chi/g_\text{eff}$  are shown for two choices of the
effective coupling:  $g_\text{eff} = 1$ (light gray) and $g_\text{eff}= 4\pi$ (dark gray).\label{fig:LimitLambdab}}
\end{figure}

Given the results of Table~\ref{tab:bkg0muWITHB} we proceed to set limits at 90\%
CL on the cutoff scale (see Table~\ref{tab:MonobLimit}) using the LHC
CL$_\mathrm{s}$ procedure. To quantify the validity of the EFT we follow the
discussion in Section~\ref{sec:EFT0mu}, considering an interaction
mediated by an $s$-channel produced particle. The operator of
Eq.~(\ref{eq:Os}) is suppressed by an additional factor
$m_\PQb$/$\Lambda$ with respect to the operators in
Eq.~(\ref{eq:OvOva}). As a result, for a given value of the coupling
$g_\text{eff}$, smaller values of $\Lambda$ are probed in this
case. The observed limit stays below the contours derived for
$R_{\Lambda} = 80\%$, even when the coupling is fixed to the largest
value considered, $g_\text{eff} = 4\pi$, as shown in the left plot
of Fig.~\ref{fig:LimitLambdab}. For the same choice of coupling,
the derived limit on $\Lambda$ would correspond to $R_{\Lambda}
\approx 25\%$, as shown in the right plot of
Fig.~\ref{fig:LimitLambdab}. Only for $g_\text{eff} > 4\pi$ does the
observed limit correspond to values of $R_{\Lambda}>80\%$. This
requirement implies a UV completion of the EFT beyond
the perturbative regime. For this reason, this result is not
interpreted in terms of an exclusion limit on $\sigma_{N\chi}$.

\begin{table}
\centering
\topcaption{\label{tab:MonobLimit} The 90\% CL limits on DM production in
  the case of scalar couplings. Here, $\sigma^\text{obs}_\mathrm{UL}$ is the observed upper limit on the production cross
section, $\Lambda^\text{obs}_\mathrm{LL}$ and $\Lambda^\text{exp}_\mathrm{LL}$
are the observed and expected cutoff energy scale lower limit, respectively.}
\begin{tabular}{*{4}{c}}
\hline
$M_\chi$  (\GeVns{}) & $\sigma^\text{obs}_\mathrm{UL}$(pb) & $\Lambda^\text{obs}_\mathrm{LL}$
                                             (\GeVns{}) & $\Lambda^\text{exp}_\mathrm{LL}$ (\GeVns{}) \mT\mB\\
\hline
0.1  &  5.4   &  43.0  &  48.2  \\
1  &  3.8   &  45.3  &  49.9  \\
10  &  6.3   &  43.2  &  48.4 \\
100  &  0.8  &  53.7  &  55.1 \\
200  &  0.7  &  47.2  &  48.3  \\
300  &  2.8  &   32.5  &  35.8  \\
400  &  2.8  &  28.3  &  30.8  \\
1000  &  1.7  &  13.2  &  13.8 \\
\hline
\end{tabular}
\end{table}

\section{Summary}\label{sec:conclusions}

A search for dark matter has been performed studying proton-proton collisions collected
with the CMS detector at the LHC at a center-of-mass energy of
8\TeV. The data correspond to an integrated luminosity of
18.8\fbinv, collected with a dedicated high-rate trigger in 2012,
made possible by the creation of parked data, and processed during the LHC shutdown
in 2013.

Events with at least two jets are analyzed by studying the
distribution in the ($\MR$, $\RR^2$) plane, in an
event topology complementary to that of monojet searches. Events with one or
two muons are used in conjunction with simulated samples, to predict
the expected background from standard model processes, mainly
$\cPZ$+jets and $\PW$+jets. The analysis is performed on events both
with and without b-tagged jets, originating from the hadronization of
a bottom quark, where in the latter case the dominant background comes from
$\ttbar$.

No significant excess is observed. The results are presented as
exclusion limits on dark matter production at 90\% confidence level
for models based on effective operators and for different assumptions
on the interaction between the dark matter particles and the colliding
partons. Dark matter production at the LHC is excluded for a mediator
mass scale $\Lambda$ below 1\TeV in the case of a vector or axial
vector operator. While the sensitivity achieved is similar to those of previously
published searches, this analysis complements those results since the
use of razor variables provides more inclusive selection criteria and
since the exploitation of parked data allows events with small values of $\MR$ to be included.

\begin{acknowledgments}
We congratulate our colleagues in the CERN accelerator departments for
the excellent performance of the LHC and thank the technical and
administrative staffs at CERN and at other CMS institutes for their
contributions to the success of the CMS effort. In addition, we
gratefully acknowledge the computing centers and personnel of the
Worldwide LHC Computing Grid for delivering so effectively the
computing infrastructure essential to our analyses. Finally, we
acknowledge the enduring support for the construction and operation of
the LHC and the CMS detector provided by the following funding
agencies: the Austrian Federal Ministry of Science, Research and
Economy and the Austrian Science Fund; the Belgian Fonds de la
Recherche Scientifique, and Fonds voor Wetenschappelijk Onderzoek; the
Brazilian Funding Agencies (CNPq, CAPES, FAPERJ, and FAPESP); the
Bulgarian Ministry of Education and Science; CERN; the Chinese Academy
of Sciences, Ministry of Science and Technology, and National Natural
Science Foundation of China; the Colombian Funding Agency
(COLCIENCIAS); the Croatian Ministry of Science, Education and Sport,
and the Croatian Science Foundation; the Research Promotion
Foundation, Cyprus; the Ministry of Education and Research, Estonian
Research Council via IUT23-4 and IUT23-6 and European Regional
Development Fund, Estonia; the Academy of Finland, Finnish Ministry of
Education and Culture, and Helsinki Institute of Physics; the Institut
National de Physique Nucl\'eaire et de Physique des Particules~/~CNRS,
and Commissariat \`a l'\'Energie Atomique et aux \'Energies
Alternatives~/~CEA, France; the Bundesministerium f\"ur Bildung und
Forschung, Deutsche Forschungsgemeinschaft, and Helmholtz-Gemeinschaft
Deutscher Forschungszentren, Germany; the General Secretariat for
Research and Technology, Greece; the National Scientific Research
Foundation, and National Innovation Office, Hungary; the Department of
Atomic Energy and the Department of Science and Technology, India; the
Institute for Studies in Theoretical Physics and Mathematics, Iran;
the Science Foundation, Ireland; the Istituto Nazionale di Fisica
Nucleare, Italy; the Ministry of Science, ICT and Future Planning, and
National Research Foundation (NRF), Republic of Korea; the Lithuanian
Academy of Sciences; the Ministry of Education, and University of
Malaya (Malaysia); the Mexican Funding Agencies (CINVESTAV, CONACYT,
SEP, and UASLP-FAI); the Ministry of Business, Innovation and
Employment, New Zealand; the Pakistan Atomic Energy Commission; the
Ministry of Science and Higher Education and the National Science
Center, Poland; the Funda\c{c}\~ao para a Ci\^encia e a Tecnologia,
Portugal; JINR, Dubna; the Ministry of Education and Science of the
Russian Federation, the Federal Agency of Atomic Energy of the Russian
Federation, Russian Academy of Sciences, and the Russian Foundation
for Basic Research; the Ministry of Education, Science and
Technological Development of Serbia; the Secretar\'{\i}a de Estado de
Investigaci\'on, Desarrollo e Innovaci\'on and Programa
Consolider-Ingenio 2010, Spain; the Swiss Funding Agencies (ETH Board,
ETH Zurich, PSI, SNF, UniZH, Canton Zurich, and SER); the Ministry of
Science and Technology, Taipei; the Thailand Center of Excellence in
Physics, the Institute for the Promotion of Teaching Science and
Technology of Thailand, Special Task Force for Activating Research and
the National Science and Technology Development Agency of Thailand;
the Scientific and Technical Research Council of Turkey, and Turkish
Atomic Energy Authority; the National Academy of Sciences of Ukraine,
and State Fund for Fundamental Researches, Ukraine; the Science and
Technology Facilities Council, UK; the US Department of Energy, and
the US National Science Foundation.

Individuals have received support from the Marie-Curie programme and the European Research Council and EPLANET (European Union); the Leventis Foundation; the A. P. Sloan Foundation; the Alexander von Humboldt Foundation; the Belgian Federal Science Policy Office; the Fonds pour la Formation \`a la Recherche dans l'Industrie et dans l'Agriculture (FRIA-Belgium); the Agentschap voor Innovatie door Wetenschap en Technologie (IWT-Belgium); the Ministry of Education, Youth and Sports (MEYS) of the Czech Republic; the Council of Science and Industrial Research, India; the HOMING PLUS programme of the Foundation for Polish Science, cofinanced from European Union, Regional Development Fund; the OPUS programme of the National Science Center (Poland); the Compagnia di San Paolo (Torino); MIUR project 20108T4XTM (Italy); the Thalis and Aristeia programmes cofinanced by EU-ESF and the Greek NSRF; the National Priorities Research Program by Qatar National Research Fund; the Rachadapisek Sompot Fund for Postdoctoral Fellowship, Chulalongkorn University (Thailand); the Chulalongkorn Academic into Its 2nd Century Project Advancement Project (Thailand); and the Welch Foundation, contract C-1845; and  the Weston Havens Foundation (USA).
\end{acknowledgments}

\bibliography{auto_generated}
\appendix

\section*{Appendix}\label{sec:appendix}

\section{Background estimation and observed yield}

In this section, we provide the background estimate and the
observed yield for each bin of the ($\MR$, $\RR^2$)
plane.

Tables~\ref{tab:LOOKUP_VL}-\ref{tab:LOOKUP_VH} show the expected and
observed yields in each $\RR^2$ bin of each $\MR$
category for the 0$\mu$ sample.  Tables~\ref{tab:LOOKUP_B} and
\ref{tab:LOOKUP_BB} show the corresponding values for the 0$\mu$b and
the 0$\mu$bb samples, respectively.

\begin{table}[hb]
\centering
\topcaption{\label{tab:LOOKUP_VL} Background estimates and observed
    yield for each $\RR^2$ bin in the VL $\MR$
    category.}
\begin{tabular}{*{5}{c}}
  \hline
  $\RR^2$ range & 0.5--0.55 &  0.55--0.6 &  0.6--0.65 & 0.65--0.7 \mT\mB\\
  \hline
  Observed & 2049 & 1607 & 1352 & 1147 \\
  Estimated & $2350\pm720$  & $1810\pm450$ & $1530\pm180$ & $1240\pm110$\\[\cmsTabSkip]
  \hline
  $\RR^2$ range & 0.7--0.75 &  0.75--0.8 &  0.8--0.85 & 0.85--0.9 \mT\mB\\
  \hline
  Observed & 1026 & 896 & 880 & 744 \\
  Estimated & $1090\pm140$ & $1081\pm76$ & $876\pm97$ & $909\pm63$ \\[\cmsTabSkip]
  \hline
  $\RR^2$ range  &  0.9--0.95  & 0.95--1.0 & \multicolumn{2}{c}{1.0--2.5} \mT\mB\\
  \hline
  Observed  & 688 & 499 &  \multicolumn{2}{c}{735}  \\
  Estimated & $674\pm67$ & $521\pm43$ &  \multicolumn{2}{c}{$694\pm62$}  \\
  \hline
\end{tabular}
\end{table}
\begin{table}
\centering
\topcaption{\label{tab:LOOKUP_L } Background estimates and observed
    yield for each $\RR^2$ bin in the L $\MR$
    category.}
\begin{tabular}{*{4}{c}}
  \hline
  $\RR^2$ range & 0.5--0.575 &  0.575--0.65 &  0.65--0.75  \mT\mB\\
  \hline
  Observed & 1088 & 765 & 682  \\
  Estimated & $1220\pm120$ & $828\pm65$ & $810\pm210$ \\[\cmsTabSkip]
  \hline
  $\RR^2$ range & 0.75--0.85 &  0.85--0.95 &  0.95--2.5  \mT\mB\\
  \hline
  Observed & 565 & 395 & 290  \\
  Estimated & $551\pm59$  & $454\pm32$ & $304\pm43$ \\
  \hline
\end{tabular}
\end{table}
\begin{table}
\centering
\topcaption{\label{tab:LOOKUP_H} Background estimates and observed
    yield for each $\RR^2$ bin in the H $\MR$
    category.}
\begin{tabular}{*{4}{c}}
  \hline
  $\RR^2$ range & 0.5--0.575 &  0.575--0.65 &  0.65--0.75  \mT\mB\\
  \hline
  Observed & 513 & 328 & 279 \mT\mB\\
  Estimated & $560\pm550$ & $330^{+360}_{-330}$ & $275\pm41$\mT\mB\\
  \hline
  $\RR^2$ range & 0.75--0.85 &  0.85--0.95 &  0.95--2.5  \mT\mB\\
  \hline
  Observed & 203 & 151 & 85   \mT\mB\\
  Estimated & $242\pm18$ & $171^{+173}_{-171}$ & $74\pm17$ \mT\mB \\
  \hline
\end{tabular}
\end{table}
\begin{table}
\centering
\topcaption{\label{tab:LOOKUP_VH} Background estimates and observed
    yield for each $\RR^2$ bin in the VH $\MR$
    category.}
\begin{tabular}{*{5}{c}}
  \hline
  $\RR^2$ range & 0.5--0.6 &  0.6--0.7 &  0.7--0.95 & 0.95--2.5 \mT\mB\\
  \hline
  Observed & 117 & 58 & 75 & 11 \\
  \hline \mT\mB
  Estimated & $100^{+150}_{-100}$ & $59\pm36$ & $75\pm30$ & $9\pm7$ \\
  \hline
\end{tabular}
\end{table}
\begin{table}
\centering
\topcaption{\label{tab:LOOKUP_B}Background estimates and observed
    yield for each bin in the $0\mu$b signal region.}
\begin{tabular}{*{5}{c}}
  \hline
  $\RR^2$ range & 0.5--0.6 &  0.6--0.75 &  0.75--0.9 & 0.9--2.5 \mT\mB\\
  \hline
  Observed & 760 & 807 & 469 & 246 \\
  Estimated & $850\pm170$ & $620\pm120$ & $470\pm110$ & $320\pm160$ \\
  \hline
\end{tabular}
\end{table}

\begin{table}
\centering
\topcaption{\label{tab:LOOKUP_BB}Background estimates and observed
    yield for each bin in the $0\mu$bb signal region.}
\begin{tabular}{*{5}{c}}
  \hline
  $\RR^2$ range & 0.5--0.6 &  0.6--0.75 &  0.75--0.9 & 0.9--2.5 \mT\mB\\
  \hline
  Observed & 122 & 80 & 31 & 14\\
  Estimated & $135\pm30$ & $81\pm18$ & $36\pm8$ & $19\pm9$ \\
  \hline
\end{tabular}
\end{table}

\cleardoublepage \section{The CMS Collaboration \label{app:collab}}\begin{sloppypar}\hyphenpenalty=5000\widowpenalty=500\clubpenalty=5000\textbf{Yerevan Physics Institute,  Yerevan,  Armenia}\\*[0pt]
V.~Khachatryan, A.M.~Sirunyan, A.~Tumasyan
\vskip\cmsinstskip
\textbf{Institut f\"{u}r Hochenergiephysik der OeAW,  Wien,  Austria}\\*[0pt]
W.~Adam, E.~Asilar, T.~Bergauer, J.~Brandstetter, E.~Brondolin, M.~Dragicevic, J.~Er\"{o}, M.~Flechl, M.~Friedl, R.~Fr\"{u}hwirth\cmsAuthorMark{1}, V.M.~Ghete, C.~Hartl, N.~H\"{o}rmann, J.~Hrubec, M.~Jeitler\cmsAuthorMark{1}, A.~K\"{o}nig, M.~Krammer\cmsAuthorMark{1}, I.~Kr\"{a}tschmer, D.~Liko, T.~Matsushita, I.~Mikulec, D.~Rabady, N.~Rad, B.~Rahbaran, H.~Rohringer, J.~Schieck\cmsAuthorMark{1}, R.~Sch\"{o}fbeck, J.~Strauss, W.~Treberer-Treberspurg, W.~Waltenberger, C.-E.~Wulz\cmsAuthorMark{1}
% ...
\vskip\cmsinstskip
% ...
1:~~Also at Vienna University of Technology, Vienna, Austria\\
% ...

\end{sloppypar}
\end{document}